\documentclass[12pt]{article}
\usepackage{amsmath}
\usepackage{amsfonts}
\usepackage{amssymb}
\usepackage{mathrsfs}
\usepackage{multicol}
\usepackage{color}
\usepackage{graphicx}
\usepackage{pstricks}
\usepackage{pst-node}
\usepackage{wrapfig}
\usepackage{enumerate}
\usepackage{caption}
\usepackage{cite}
\usepackage{amsfonts,amssymb}
\usepackage{amsthm}
\usepackage{amscd}
\usepackage{caption}
\usepackage{subcaption}

\definecolor{gray1}{gray}{0.1}
\definecolor{gray2}{gray}{0.2}
\definecolor{gray3}{gray}{0.3}
\definecolor{gray4}{gray}{0.4}
\definecolor{gray5}{gray}{0.5}
\definecolor{gray6}{gray}{0.6}
\definecolor{gray7}{gray}{0.7}
\definecolor{gray8}{gray}{0.8}
\definecolor{gray9}{gray}{0.9}

\newpsobject{showgrid}{psgrid}{subgriddiv=1,griddots=10,gridlabels=6pt}

\definecolor{dark-green}{rgb}{0,0.7,0}
\definecolor{dark-blue}{rgb}{0,0.2,0.5}
\definecolor{med-blue}{rgb}{0,0.7,1}
\definecolor{mblue}{rgb}{0,0.2,1}
\definecolor{cnc}{rgb}{0.8,0,0}
\definecolor{light-red}{rgb}{1,0.8,0.8}
\definecolor{dark-yelow}{rgb}{1,0.8,0}
\definecolor{light-blue}{rgb}{0.8,0.9,1}
\definecolor{verylight-blue}{rgb}{0.93,0.95,1}
\definecolor{light-yelow}{rgb}{1,0.9,0.8}
\definecolor{grey}{gray}{0.88}

\newpsobject{showgrid}{psgrid}{subgriddiv=1,griddots=10,gridlabels=6pt}

\newcommand{\be}{\begin{equation}}
\newcommand{\ee}{\end{equation}}
\newcommand{\bea}{\begin{eqnarray}}
\newcommand{\eea}{\end{eqnarray}}
\newcommand{\beann}{\begin{eqnarray*}}
\newcommand{\eeann}{\end{eqnarray*}}
\def\IZ{\rlx\hbox{\sf Z\kern-.4em Z}}
\def\IR{\rlx\hbox{\rm I\kern-.18em R}}
\def\IC{\rlx\hbox{\,$\inbar\kern-.3em{\rm C}$}}
\def\one{\hbox{{1}\kern-.25em\hbox{l}}}
\newcommand{\csch}{\textrm{csch}}

\DeclareMathOperator{\sech}{sech}

\begin{document}

\thispagestyle{empty}

\setlength{\abovecaptionskip}{10pt}

\begin{center}
{\Large\bfseries\sffamily{Some novel considerations about the collective coordinates approximation for the scattering of $\phi^4$ kinks}}
\end{center}
\vskip 3mm
\begin{center}
{\bfseries{\sffamily{Carlos F.S. Pereira$^{\rm 1}$, Gabriel Luchini$^{\rm 1}$, Tadeu Tassis$^{\rm 2}$,Clisthenis P. Constantinidis$^{\rm 1}$}}}\\
\vskip 3mm{
$^{\rm 1}$\normalsize Departamento de F\'isica,
Universidade Federal do Esp\'irito Santo (UFES),\\
CEP 29075-900, Vit\'oria-ES, Brazil}
\vskip 3mm{
$^{\rm 2}$\normalsize Departamento de F\'isica,
Universidade Federal do ABC (UFABC),\\
CEP 09210-580,- Bang\'u, Santo Andr\'e-SP, Brazil}
\end{center}

\begin{abstract} 
The collective coordinates approximation for the kink/anti-kink scattering in the $1+1$ dimensional $\phi^4$ model is considered and we discuss how the results found in the current literature on the topic can be improved by giving the analytical expression for the lagrangian in the space of parameters. A comprehensive discussion of the role of the collective coordinates approximation in this particular situation is given for completeness.
\end{abstract}

\section{Introduction}
In this paper we revisit the problem of \textit{kink}/\textit{anti}-\textit{kink} scattering in the $\phi^4$ model using collective coordinates. Although this subjet has been extensively discussed recently in the literature \cite{Weigel:2013kwa, Takyi:2016tnc, Kevrekidis:2019zon} and also models which are constructed from the $\phi^4$ model are also highly considered in this context \cite{Adam_2019, adam2019kinkantikink,adam2019thick},  our results  bring some improvements  with respect to the previous approaches to find the effective theory based on collective coordinates.

Collective coordinates can be used as an approximation method for studying scattering processes of solitons in non-integrable field theories, such as for the $\phi^4$ model. It appears as an alternative for the construction of multisoliton configurations by reducing the degrees of freedom which characterize the dynamics of these models by essentially making some suitable choice of a set of relevant dynamical  variables (the collective coordinates) for the problem whose evolution in time will describe the important features of the scattering.

In the $\phi^4$ model \textit{kink}/\textit{anti}-\textit{kink} scattering, the full numerical evolution of this configuration shows a resonanse behaviour of the system where the individual solitons will form a bounded system during a certain time interval which is then undone and the solitons are again free to move\cite{Peyrard:1984qn,Campbell:1983xu, Campbell:1986mg, Anninos:1991un, moshir}. 

The collective coordinates approximation has been used in attempts to understand not only the scattering of these solitons as well as a way to shed some light on this intriguing resonanse phenomenon. Important contributions were made\cite{Sugiyama:1979mi} pointing to the fact that during the collision the excitation mode of the $\phi^4$ solitons becomes important. This is corroborated essentially by the fact that the collective coordinate approximation is considerably improved and much closer to the full simuations results when the possibility of this excitation is considered.

One of the drawbacks of this approximation method is a technical issue: the construction of the lagrangian in the space of parameters requires the calculation of many highly non-trivial integrals. In the literature several mathematical methods have been employed and they are generally quite complicate. Perhaps, this difficulty has been forcing many authors to find arguments to ignore some of those integrals in their approximation. The main goal of this paper is to present the full lagrangian which describes the dynamics of the \textit{kink}/\textit{anti}-\textit{kink} scattering in the $\phi^4$ model for the collective coordinates which encode information about the translation of the solitons and their excitation due to mutual interaction. 


The first sections of the paper are devoted to an introduction of the system we are going to study as a way to fix notation and ideas. In section \ref{1}  we discuss the topological aspects of the $\phi^4$ model and some characteristics of the dynamics of the \textit{kink} and \textit{anti}-\textit{kink} solutions. Collective coordinates are introduced in section \ref{2} where the method is illustrated for a simple case of a single (\textit{anti})-\textit{kink} solution. Then we discuss how this approximation is implemented in the case of the scattering of these objects by looking at was is known in the literature and showing how it can be improved with the method of integration which is presented in one of the appendices. Finally, in section \ref{3}, we draw our conclusions on the role of the collective coordinates as employed here and our perspectives concerning the method.

\section{The model} \label{1}
\begin{figure}[h!]
  \centering
    \includegraphics[width=0.59\textwidth]{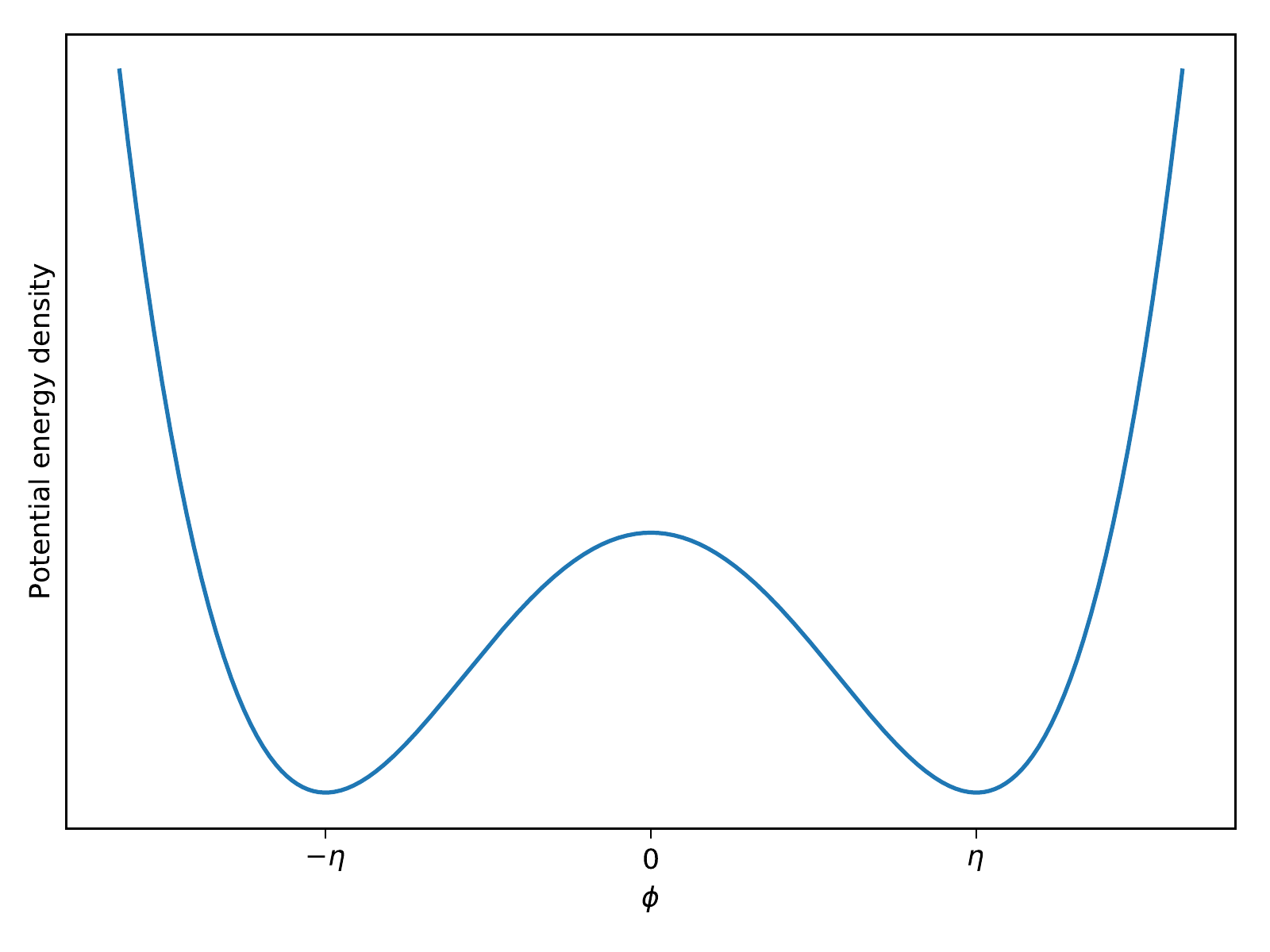}
    \caption{The potential energy density has two degenerate vacua at $\phi = \pm \eta$.} 
    \label{fig: pot_dens}
\end{figure}

The so called $\phi^4$ model has a wide range of applications\cite{Campbell:1986mg}, from condensed matter physics\cite{bishop} to high energy theories\cite{manton2004topological}, cosmology\cite{vilen,vachas}, biological systems\cite{Wittkowski_2014} and so on. Our interest here is on the $\phi^4$ model for which the field $\phi$ is a real Lorentz scalar in $1+1$ dimensional Minkowski space-time with metric $g_{\mu\nu}={\rm diag}(1,-1)$. The action for this model is defined as
\be
\label{eq: action phi4}
\mathcal{S} = \int d^2x\; \left(\frac{1}{2}\partial_\mu \phi \partial^\mu \phi - \mathcal{U}(\phi)\right),
\ee
where the energy density shown if figure \ref{fig: pot_dens} is
\be 
\label{eq: potential phi4}
\mathcal{U}(\phi)=\frac{\lambda}{4}\left(\phi^2-\eta^2\right)^2,
\ee
with $\lambda \geq 0$ for boundedness of the potential whose vacua are at $\phi = \pm \eta \in \mathbb{R}$.

The dynamical equation of the field
\be
\label{eq: eom phi4}
\partial_\mu \partial^\mu \phi + \lambda\left(\phi^2-\eta^2\right)\phi=0,
\ee
is highly non-linear and can be approached by two different directions. First there are solutions of the form $\phi = \pm \eta + \delta \phi + \mathcal{O}\left((\delta\phi)^2\right)$, with $\delta \phi \ll \eta$ a linear perturbation of a vacuum state, which consist of the trivial solution of the model, i.e., a solution with vanishing energy; the space of these solutions is known as the perturbative sector. Then, there are also the so called \textit{kink} and \textit{anti-kink} configurations which appear as solutions that interpolate between the vacua of the potential energy density, definining the non-perturbative sector of the theory. They are topological solitons\cite{manton2004topological}: their existence and stability are related to the behaviour of the field at the spatial border (the topological data), so defined that the configuration has, in this case, the least possible finite energy. This interplay between the topological data and energy reveals, for this model, that the time-independent (\textit{anti-})\textit{kink} configurations are BPS solutions\cite{Bogomolny:1975de, Sutcliffe:1997ec, Adam:2013hza}, i.e. static configurations $\phi(x)$ which render stationary the static energy functional 
\be 
\label{eq: static engy}
M = \int_{-\infty}^{+\infty}\left(\frac{1}{2}\left(\frac{d\phi}{dx}\right)^2+\mathcal{U}\right)\;dx
\ee
with fixed (and constant) field values at spatial infinity, $\phi(\pm\infty)=\pm \eta$, which satisfy the first order (BPS) equation
\be
\label{eq: bps phi4}
\frac{d\phi}{dx}=\pm \sqrt{2\mathcal{U}}
\ee
and have energy proportional to the absolute value of the topological degree
\be
\label{eq: top degree}
N = \frac{1}{2\eta} \int_{-\infty}^{+\infty} \frac{d\phi}{dx}\;dx = \frac{\phi(+\infty)-\phi(-\infty)}{2\eta} = \pm 1.
\ee

\begin{figure}[h!]
  \centering
    \includegraphics[width=0.49\textwidth]{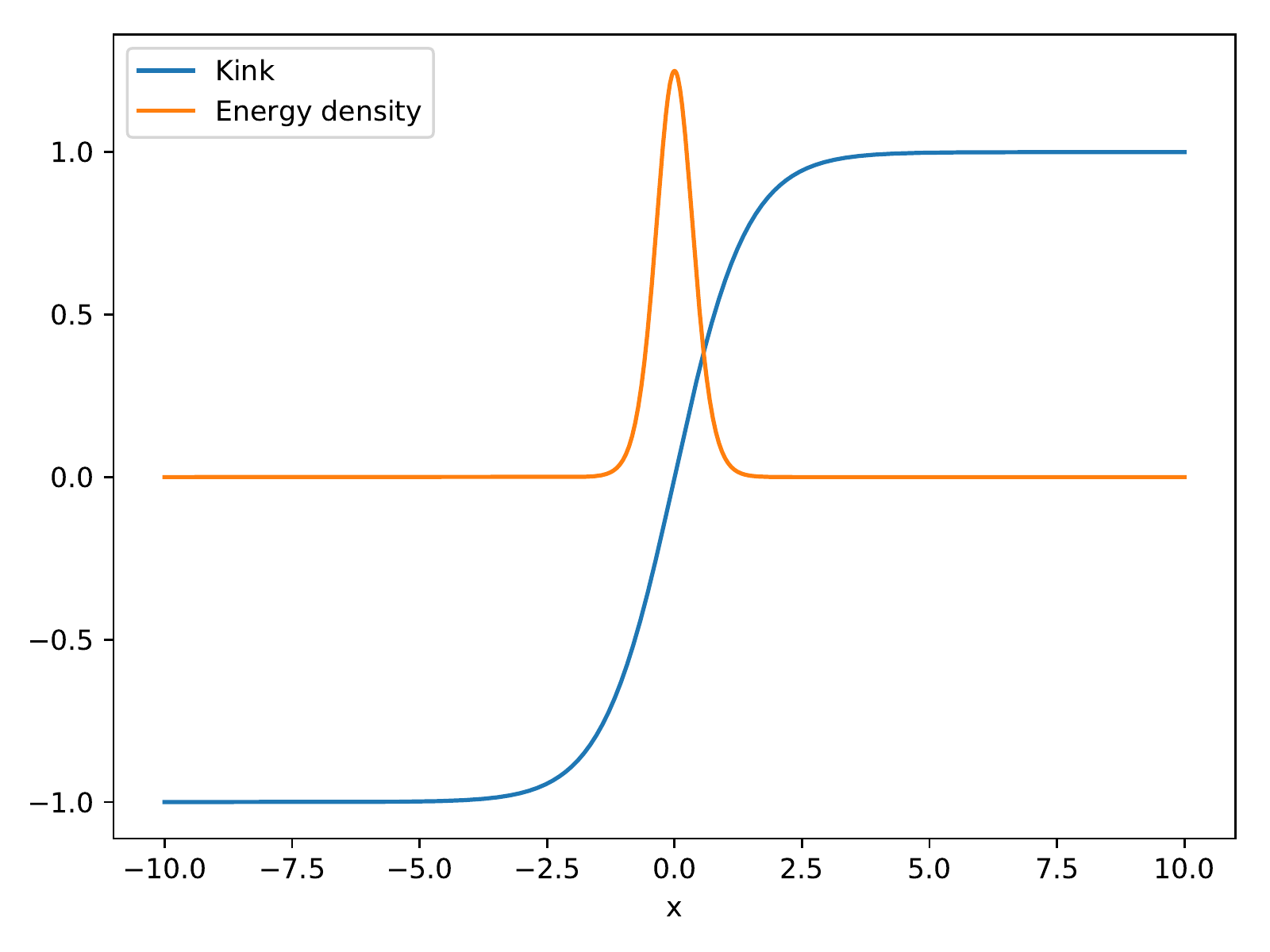}
    \caption{The energy density of the \textit{kink} configuration is a localised function around the point $x=a$ of the solution, here taken to be zero.} 
    \label{fig: kink engy}
\end{figure}

The solutions of the BPS equation (\ref{eq: bps phi4}) with the potential given in (\ref{eq: potential phi4}) can be easily obtained by direct integration and read
\be
\label{eq: sol phi4 bps}
\phi_K(x)= \eta \tanh{\left(\sqrt{\frac{\lambda}{2}}\;\eta(x-a)\right)}, \qquad \phi_{\bar{K}}(x)= -\eta \tanh{\left(\sqrt{\frac{\lambda}{2}}\;\eta(x-a)\right)},
\ee
respectively the \textit{kink} with topological degree $N = +1$ and the \textit{anti-kink} with  $N = -1$. Here, $a$ is an integration constant which defines the position of the \textit{(anti-)kink}, as it is interpreted as a relativistic particle-like\footnote{But not pointwise.} object with the static energy $M$ regarded as its rest mass. The \textit{kink} solution as well as its energy density are shown in figure \ref{fig: kink engy}. 

These BPS solutions $\phi_{K,\bar{K}}$ are stable against linear perturbations (see appendix \ref{sec: app pert}) 
\begin{equation}
\phi_{K,\bar{K}} \rightarrow \phi_{K,\bar{K}} + \epsilon(t,x).
\end{equation}

\begin{figure}[h!]
  \centering
    \includegraphics[width=0.49\textwidth]{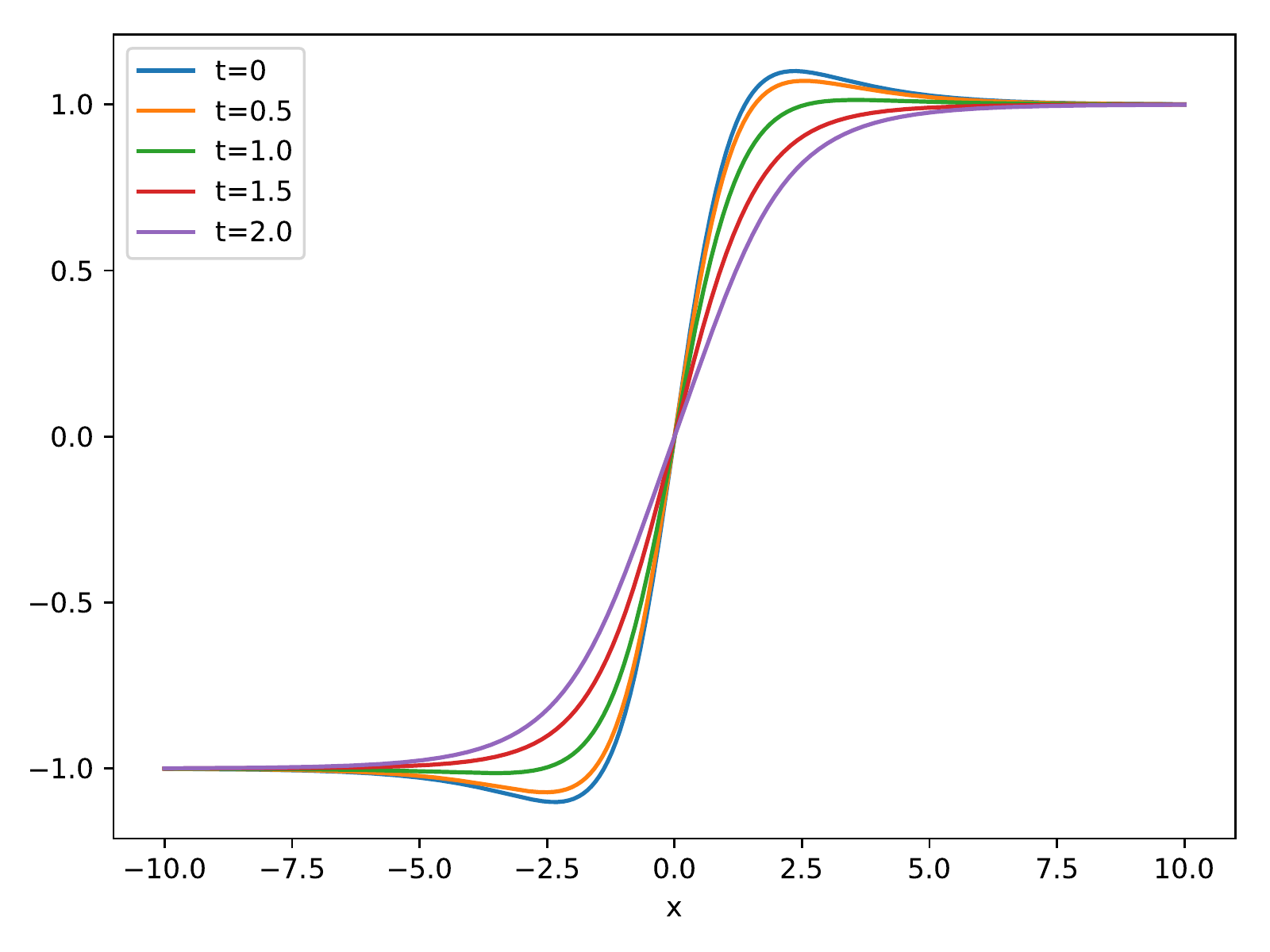}
    \caption{The (\textit{anti})-\textit{kink} solition is stable agains perturbations. Here the effect of the excitation mode on the solution is presented at different times.} 
    \label{fig: kink mode}
\end{figure}

At this level the (\textit{anti}-)\textit{kink} has a vibrational ``zero mode'', which consists of a spatial translation of the configuration along its tangent direction, thus costing no energy to the system, and a first excited mode of the form
\begin{equation}
\label{eq: mode}
\epsilon(t,x) = \frac{1}{2}\sech{(\sigma x)}\tanh{(\sigma x)}\cos\left(\sqrt{3}\sigma{t}\right), 
\end{equation}
with $\sigma = \sqrt{\frac{\lambda}{2}}\eta$. Such excitation on the \textit{kink} solution is shown in figure \ref{fig: kink mode}.

\begin{figure}[h!]
  \centering
    \includegraphics[width=0.49\textwidth]{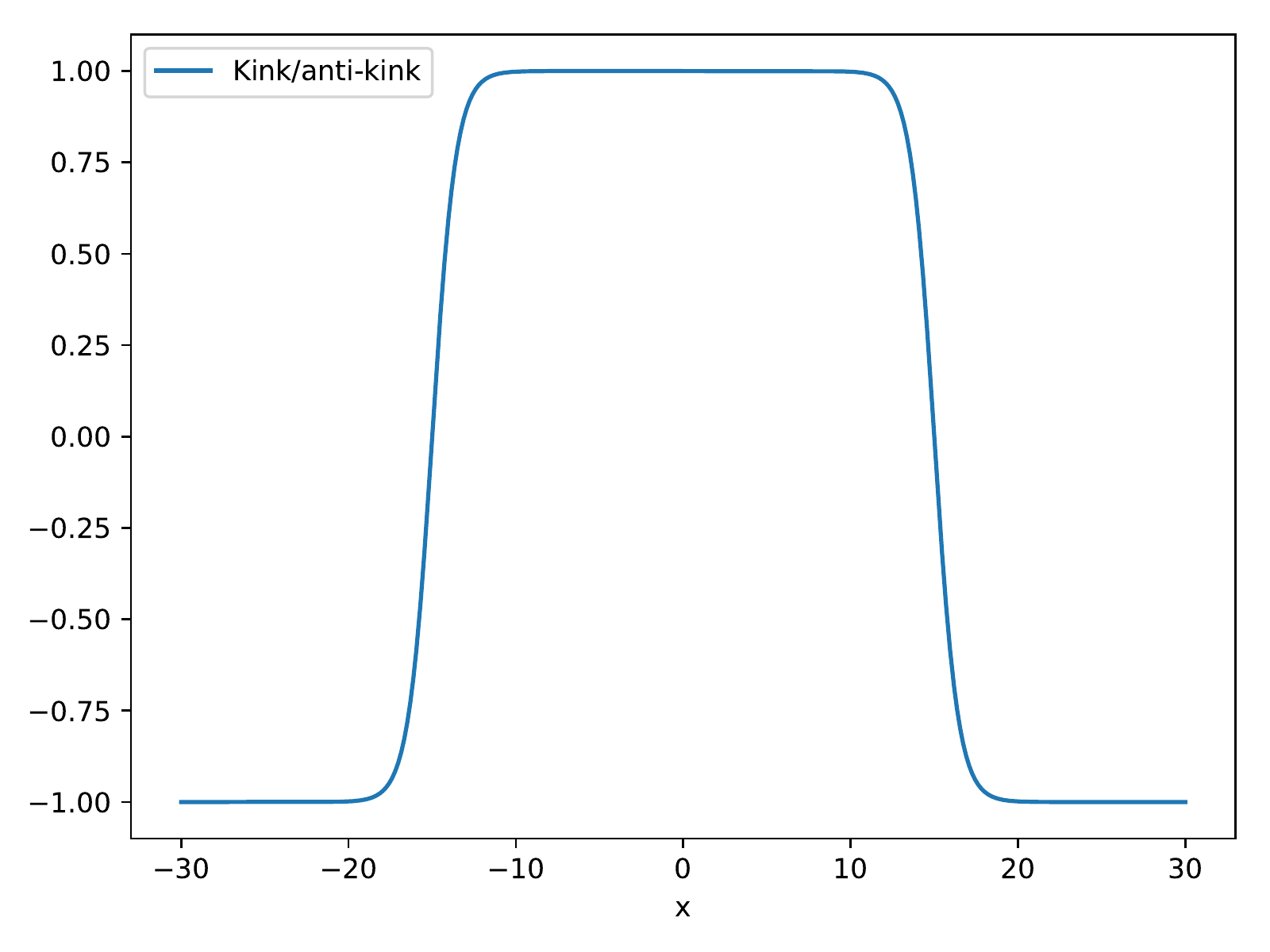}
    \caption{The \textit{kink}/\textit{anti}-\textit{kink} configuration is defined by sewing both solutions separated apart by a distance $2a$ where here, $a=15$.} 
    \label{fig: kink antikink}
\end{figure}

 The non-linear character of the model is what drives the interaction between these particle-like objects\cite{Manton_2008}. Although one cannot have a solution of the theory consisting of a moving \textit{kink}/\textit{anti-kink} pair, such a static configuration with the \textit{kink} at $x=-a$ and the \textit{anti}-\textit{kink} at $x=a$  can be constructed as shown if figure \ref{fig: kink antikink} where 
\be
\label{eq: kak config}
\phi(x) = \phi_K(\sigma(x+a))+\phi_{\bar{K}}(\sigma(x-a))-\eta.
\ee
Setting the distance $2a$ to be large enough\footnote{Which means larger than the individual length of the solitons which can be estimated as $\frac{2\sqrt{2}}{\eta}$.} this can be within some approximation considered as a legit solution. 

Looking at the well separated \textit{kink} and \textit{anti}-\textit{kink} as individual objects one can calculate an intrinsic force\cite{manton2004topological} between them given by
\begin{equation}
F = 32{\sigma}^2{\eta}^2e^{-4\sigma{a}},
\end{equation}
i.e., a weak but non-vanishing attraction. 

\begin{figure}[h!]
  \centering
    \includegraphics[width=0.49\textwidth]{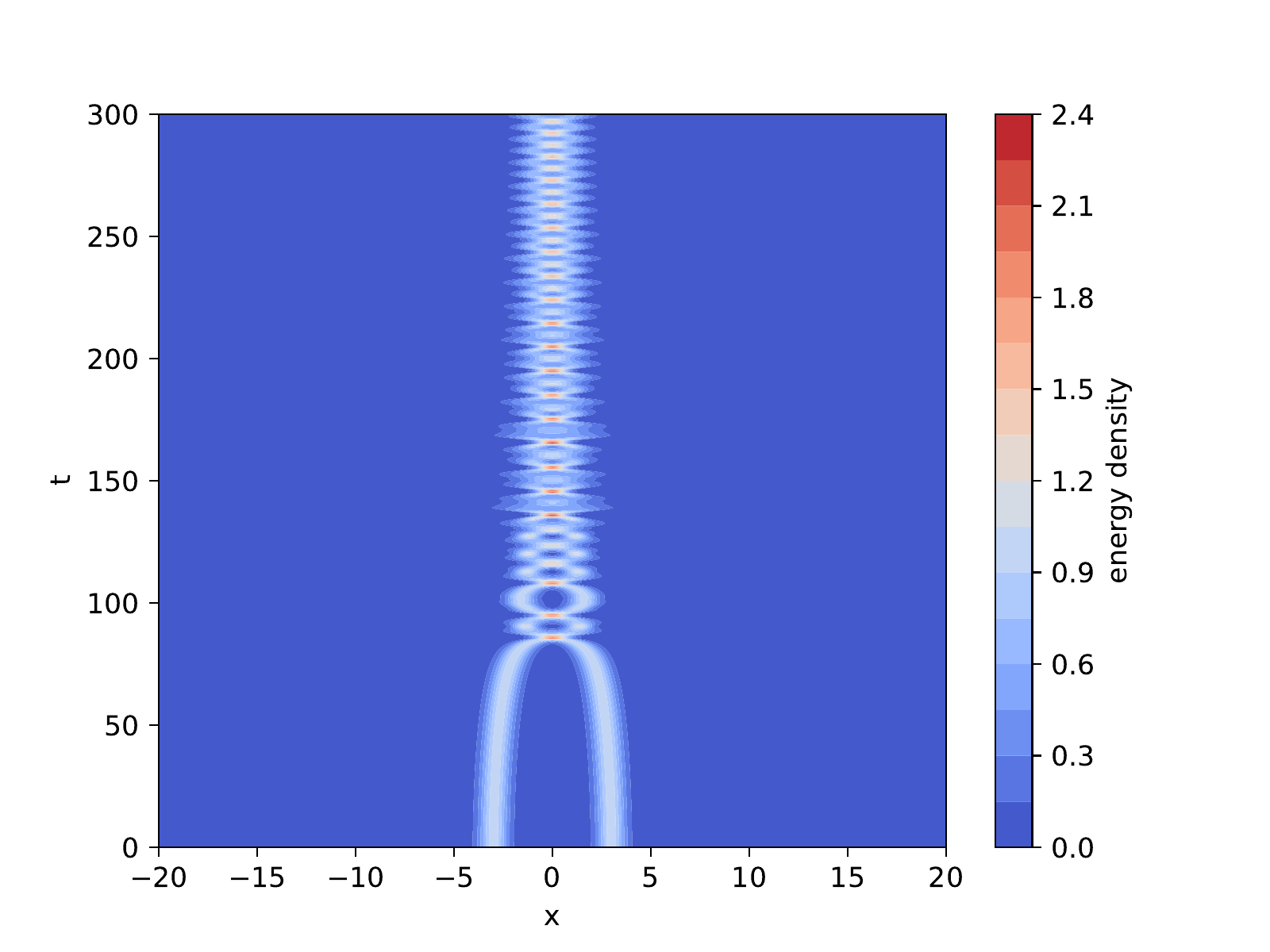}
    \caption{The energy density of the configuration is shown. We start with the static configuration (\ref{eq: kak config}) with $a = 3$ and $\eta = \sigma = 1$. The configuration will evolve numerically in time. While at first the \textit{kink} and \textit{anti}-\textit{kink} where clearly standing still at their original positions, given by the localised peaks of energy density, their mutual attraction will lead them to accelerate towards each other which can be seen after around $t= 75$.} 
    \label{fig: kak attraction}
\end{figure}

If left for long enough time this configuration, with a vanishing topological degree, is unstable against the perturbations provoked by the intrinsic force between \textit{kink} and \textit{anti-kink} and decays after a while, as shown in figure \ref{fig: kak attraction}. 

The \textit{kink}/\textit{anti-kink} pair exhibit interesting behaviour when their scattering is considered, i.e., when these individual objects acquire a relative velocity between them. Being a relativistic system, the moving (\textit{anti}-)\textit{kink} can be found by performing a Lorentz boost of the static solution:
\begin{equation}
\label{eq: mov kink}
\phi_{K,\bar{K}}(t,x) = \pm \eta \tanh{\left(\sqrt{\frac{\lambda}{2}}\;\eta\frac{(x- a - vt)}{\sqrt{1-v^2}}\right)},
\end{equation}
the parameter $v \in (-1,1)$ standing for its boost velocity.

\begin{figure}[h!]
  \centering
  \begin{subfigure}{0.49\textwidth} 
    \includegraphics[width=\textwidth]{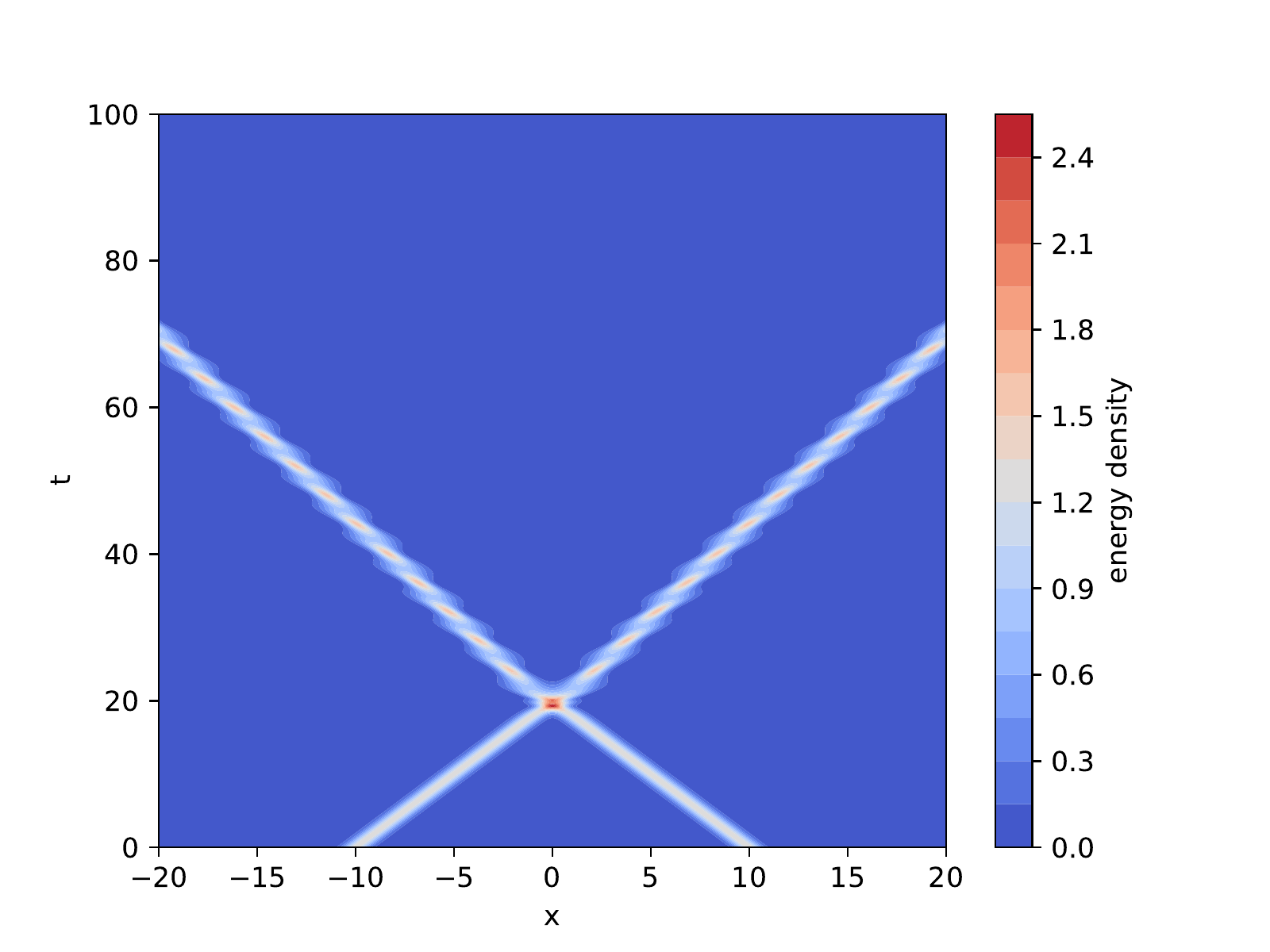}
    \caption{An elastic scattering is observed when the \textit{kink} and \textit{anti}-\textit{kink} are thrown towards each other with $\vert v \vert =0.5$ each.} 
    \label{fig: kak v04}
  \end{subfigure}
  \vspace{1em} 
  \begin{subfigure}{0.49\textwidth} 
    \includegraphics[width=\textwidth]{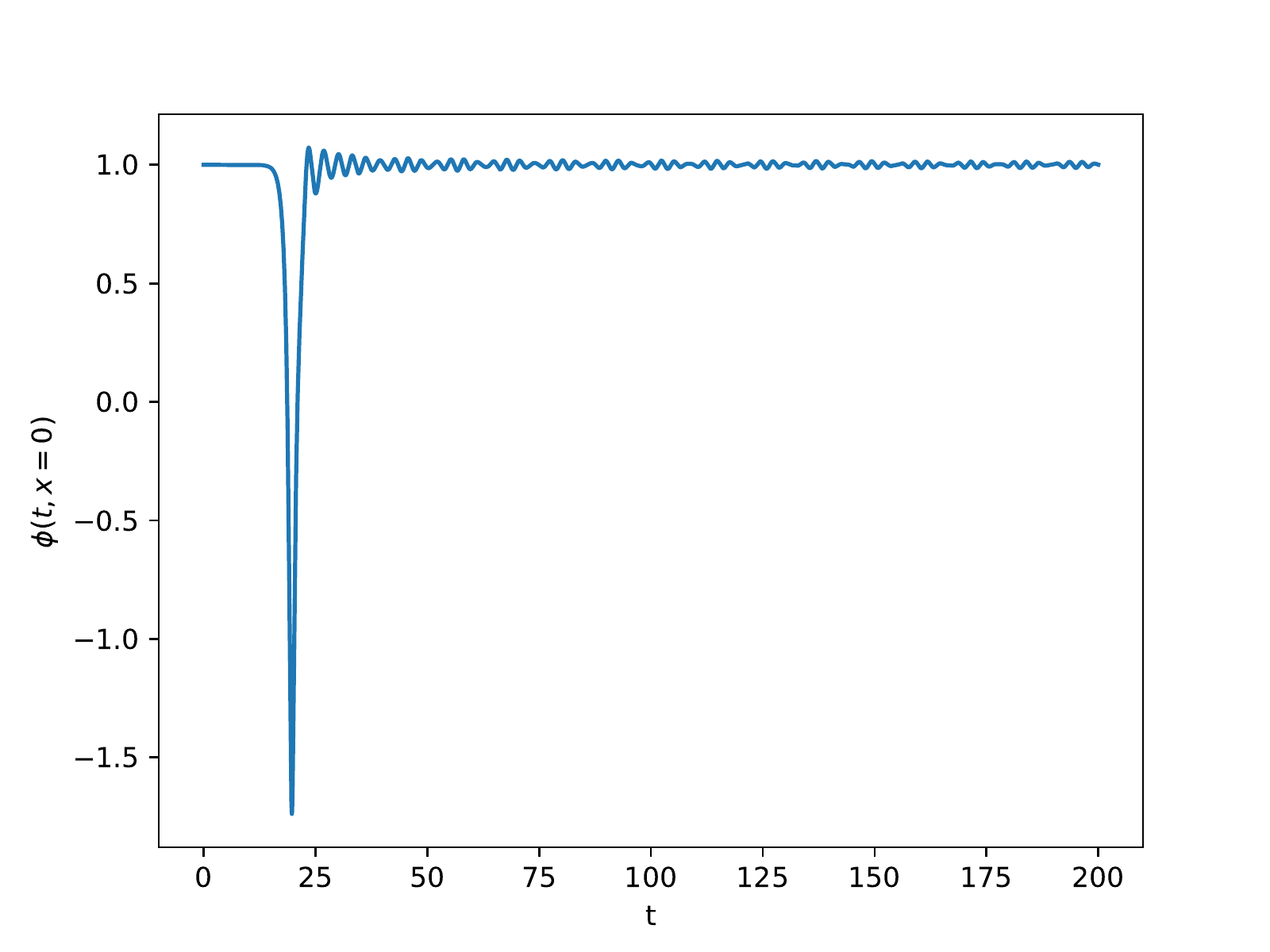}
    \caption{The value of the field $\phi$ at $x=0$ is shown for different times during the scattering process. A change occurs close to $t=20$, when the collision takes place.} 
    \label{fig: kak decay}
  \end{subfigure}
  \caption{When the solitonic objects are well separated, the field at $x=0$ has a constant value $\phi = 1$. During the collision process this value abruptly changes and in the present case, being the collision elastic, it recovers its original value. Nevertheless one observes some small wrinkles both in the value of the field and in the energy density.}
  \label{fig: elastic v04}
\end{figure}

For large values of relative velocity the topological solitons will collide elastically, thus exhibiting a behaviour similar to that of usual point particles, as seen in figure \ref{fig: elastic v04}. 

\begin{figure}[h!]
  \centering
  \begin{subfigure}{0.49\textwidth} 
    \includegraphics[width=\textwidth]{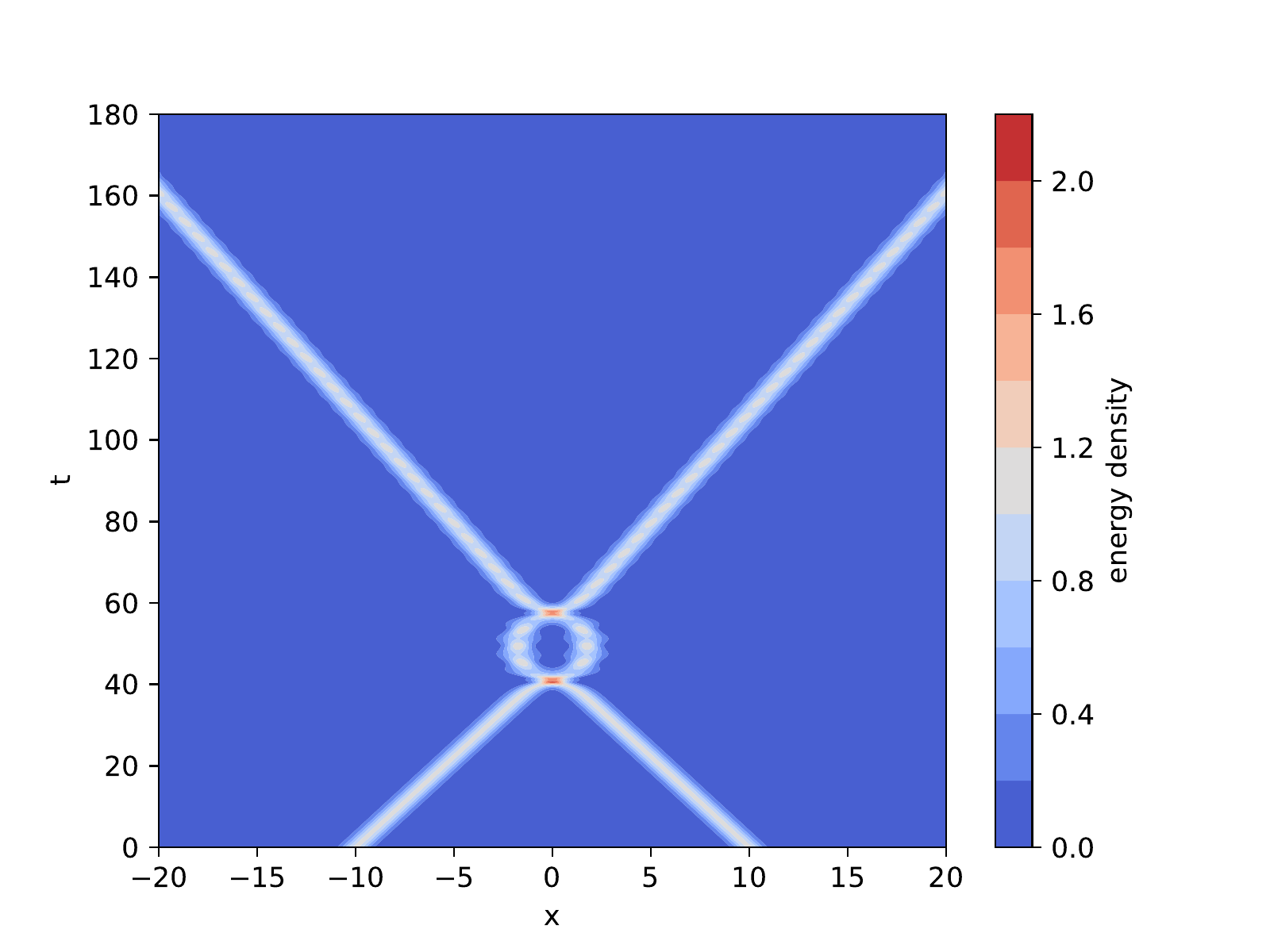}
    \caption{A bounce window is observed when the \textit{kink} and \textit{anti}-\textit{kink} are thrown towards each other with $\vert v \vert =0.2249$ each.} 
    \label{fig: kak bounce v04}
  \end{subfigure}
  \vspace{1em} 
  \begin{subfigure}{0.49\textwidth} 
    \includegraphics[width=\textwidth]{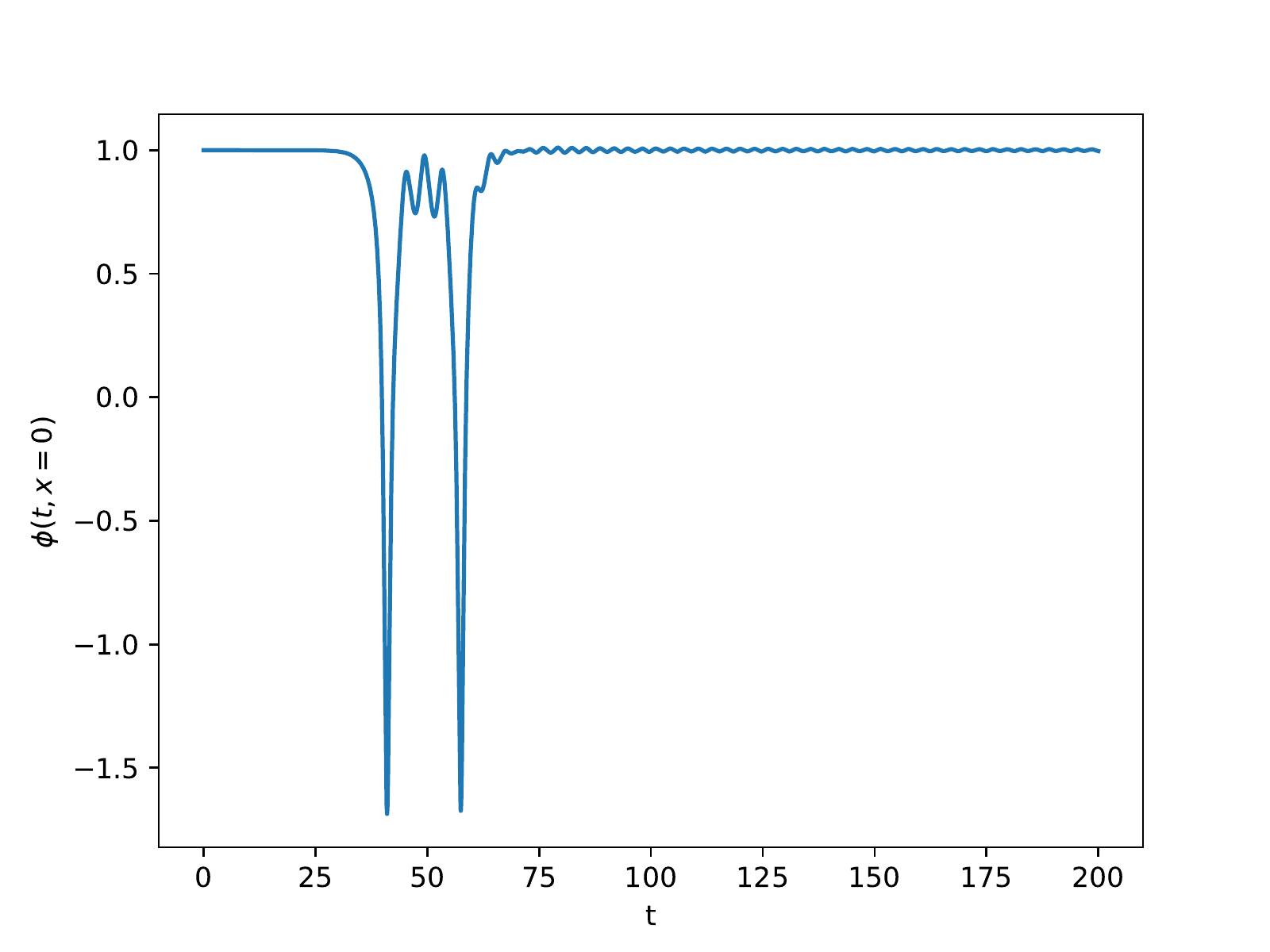}
    \caption{The value of the field $\phi$ at $x=0$ is shown for different times during the scattering process. Two changes occur, indicating two different collisions: a first one close to $t=40$ and another one close to $t=60$.} 
    \label{fig: kak field zero v04}
  \end{subfigure}
  \caption{For low values of the relative velocity between \textit{kink} and \textit{anti}-\textit{kink} one observes a resonance phenomenon where the pair of solitonic objects will form a temporary bound state after which they recover their original ``free motion''.}
  \label{fig: bounce}
\end{figure}

For some lower velocity values one sees what's called \textit{bounce windows}\cite{Sugiyama:1979mi,Takyi:2016tnc,belova,goodma}: a bound state is formed during some time lapse after which the solitonic objects recover their unbounded motion. This is shown in figure \ref{fig: bounce}.

\begin{figure}[h!]
  \centering
  \begin{subfigure}{0.49\textwidth} 
    \includegraphics[width=\textwidth]{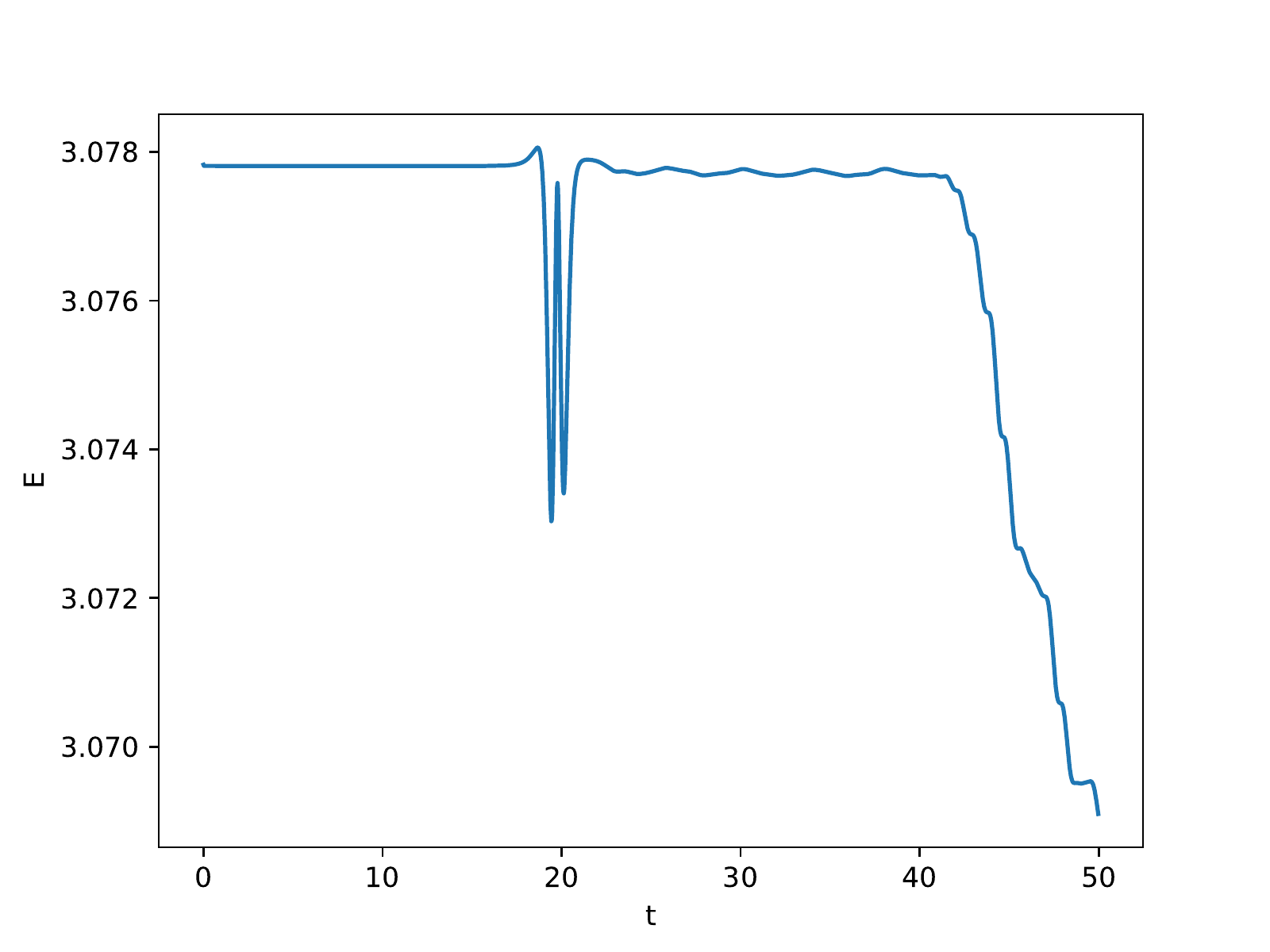}
    \caption{The energy as a function of time for the collision between \textit{kink} and \textit{anti}-\textit{kink} with initial velocity of $\vert v \vert = 0.5$ each. The observed difference between the energy after the collision and before is $\Delta E = 0.00877$} 
    \label{fig: v05 engy time}
  \end{subfigure}
  \vspace{1em} 
  \begin{subfigure}{0.49\textwidth} 
    \includegraphics[width=\textwidth]{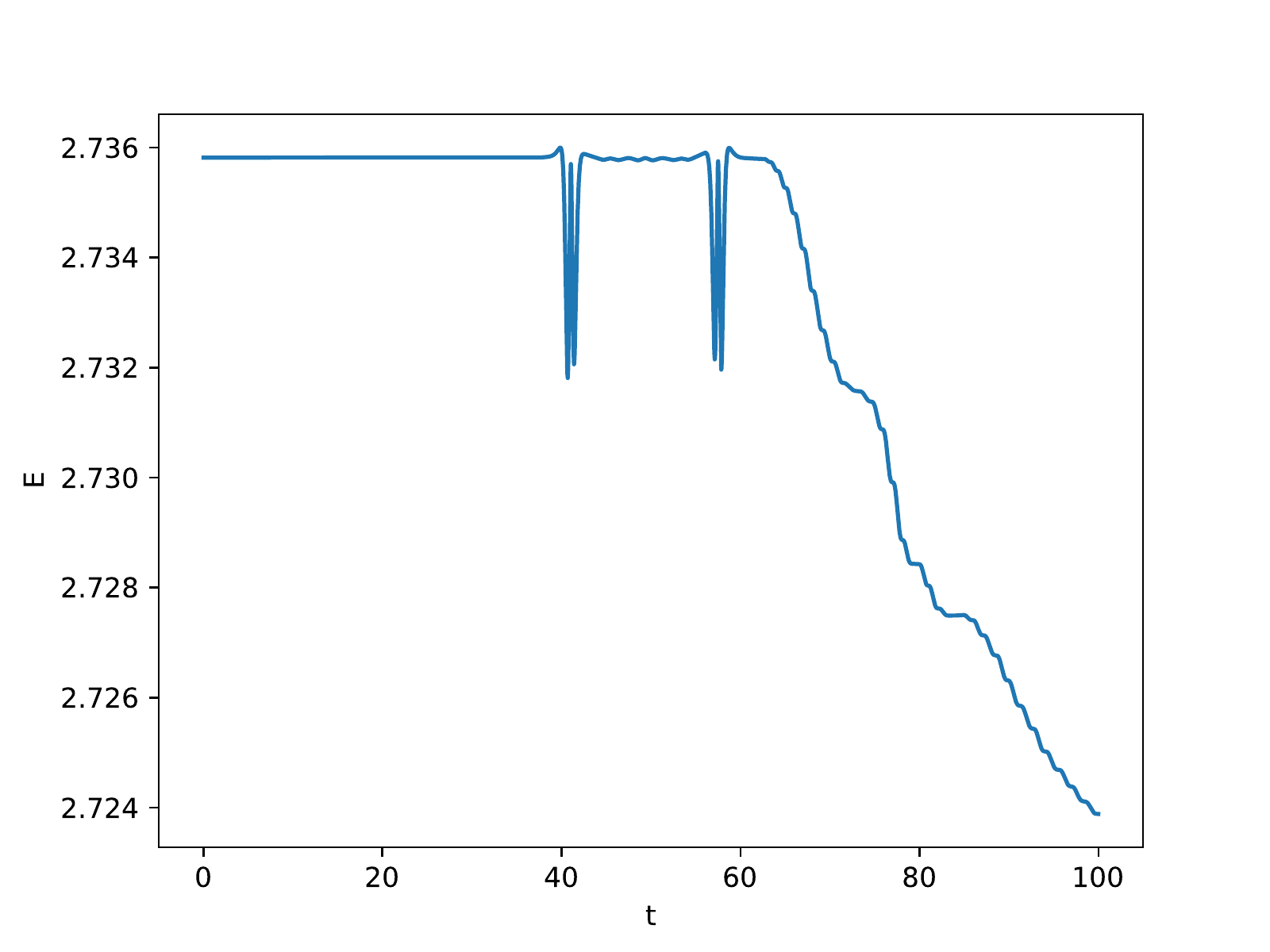}
    \caption{The energy as a function of time for the collision between \textit{kink} and \textit{anti}-\textit{kink} with initial velocity of $v = 0.2249$ each. The observed difference between the energy after the collision and before is $\Delta E = 0.00005$ } 
    \label{fig: v02249 engy time}
  \end{subfigure}
  \caption{The energy of the system is calculated in the region between $x=-20$ and $x=20$, from when the solitons start their motion towards each other, during the collision and after that for a while. It is observed that the amount of radiation dispersed in this process is very little.}
  \label{fig: engy loss}
\end{figure}

Although this model is non-integrable and therefore energy is expect to be lost in the collision process, by performing the numerical calculation of the total energy while the solitons scatter we observe that the emmition of radiation is in fact very little, as shown in figure \ref{fig: engy loss}.

This fact seems to indicate that the ripples observed in the field after the scattering are not a consequence of this process of radiation emmition only but perhaps much more related to a reorganisation of the distribution of energy within the system.

Thus, if from one hand these objects have many particle-like features, from the other they also exhibit this intriguing resonance phenomenon, which is not observed for pointwise particles.  This is interesting enough to motivate an attempt to understand which mechanism is responsible to the formation of these so called bounce windows.

\section{The collective coordinates approach}\label{2}

One way of studying the dynamics of the soliton collisions, especially when low values of relative velocity are considered, is through the so called collective coordinates method\cite{gabi,goodma, Sugiyama:1979mi, belova,Takyi:2016tnc}: the degrees of freedom of the field are encoded into time dependent parameters (the collective coordinates) whose dynamics will determine the relative motion of the solitons through a geodesic in the space where these parameters are defined as coordinates.

To study the motion of a free (\textit{anti}-)\textit{kink} moving at low velocity the collective coordinate $a(t)$ can be introduced as the position function of that object by writing
\begin{equation}
\phi = \pm \eta \tanh{\left(\sigma\left(x-a(t)\right)\right)}, \qquad \sigma = \sqrt{\frac{\lambda}{2}}\eta.
\end{equation}
By doing so we have that $\frac{\partial \phi}{\partial t} = \dot{a}\frac{\partial \phi}{\partial x}$ and all time dependence gets factorized in terms of $\dot{a}(t)$ and the lagrangian density in (\ref{eq: action phi4}) can be integrated over the spatial coordinate giving
 
\begin{equation}
\label{eq: lag coll a}
L = \int_{-\infty}^{+\infty}\mathcal{L}\,dx = \frac{1}{2}g(a)\dot{a}^2 - V(a)
\end{equation}
where, in this case, $g(a)$ is in fact the constant $g = \frac{2\eta^2}{3}\sigma$ as well as $V(a)$ is the constant function $V = \frac{2\eta^2\sigma}{3}+\frac{\lambda \eta^4}{3\sigma}$. The dynamical equations for the field are now reduced to the Euler-Lagrange equation for the collective coordinate $a(t)$, which reads $\ddot{a}=0$, whose solution is
$$
a(t) = a_0+vt,
$$
with $a_0$ and $v$ integration constants. As the result we have that the soliton will move with a constant velocity $v$ and its profile reads
\begin{equation}
\phi = \pm \eta \tanh{\left(\sigma\left(x- a_0 - vt\right)\right)}, 
\end{equation}
which approximately describes the time dependent (\textit{anti-})\textit{kink} solution 
\begin{equation}
\phi= \pm \eta \tanh{\left(\sqrt{\frac{\lambda}{2}}\;\eta\frac{x-a_0 -vt}{\sqrt{1-v^2}}\right)},
\end{equation}
for $v\ll 1$.

The description of the \textit{kink}/\textit{anti-kink} collision dynamics through collective coordinates is not as straightforward as the case for the free soliton. Indeed, by setting\cite{Sugiyama:1979mi}
\begin{equation}
\phi(t,x) = \phi_K(x+a(t)) + \phi_{\bar{K}}(x-a(t)) - \eta 
\end{equation}
the translation coordinate $a(t)$ now describes half of the relative distance between the \textit{kink} and the \textit{anti-kink} and the lagrangian obtained for $a(t)$ in the form of (\ref{eq: lag coll a}) has the functions $g(a)$ and $V(a)$ given by\footnote{The integration method which is used here is shown in details in appendix \ref{sec: appendix integrals}}
\begin{eqnarray}
g\left(a\right)&=& {{\eta}^2{\sigma}^2}\left[\frac{8}{{3}{\sigma}} -\frac{8}{\sigma}\csch^2\left({2}{\sigma}{a}\right)+{16}{a}{\coth\left({2}{\sigma}{a}\right)}{\csch^2\left({2}{\sigma}{a}\right)}\right]\\
V\left(a\right)&=&8\sqrt{2}{\sigma}^2{\eta}^2\left[{-\frac{2}{3}}+{2\sigma{a}}+{\frac{3}{\tanh\left(2{\sigma}a\right)}}\right] \nonumber\\
&+&8\sqrt{2}{\eta}^2{\sigma}^2\left[-{\frac{\left(2+6{\sigma}a\right)}{\tanh^2\left(2{\sigma}a\right)}}+{\frac{4{\sigma}a}{\tanh^3\left(2{\sigma}a\right)}}\right]
\end{eqnarray}
whose behaviour can be seen in the figure \ref{fig: kak gv} below.
\begin{figure}[h!]
\centering
\includegraphics[scale=0.45]{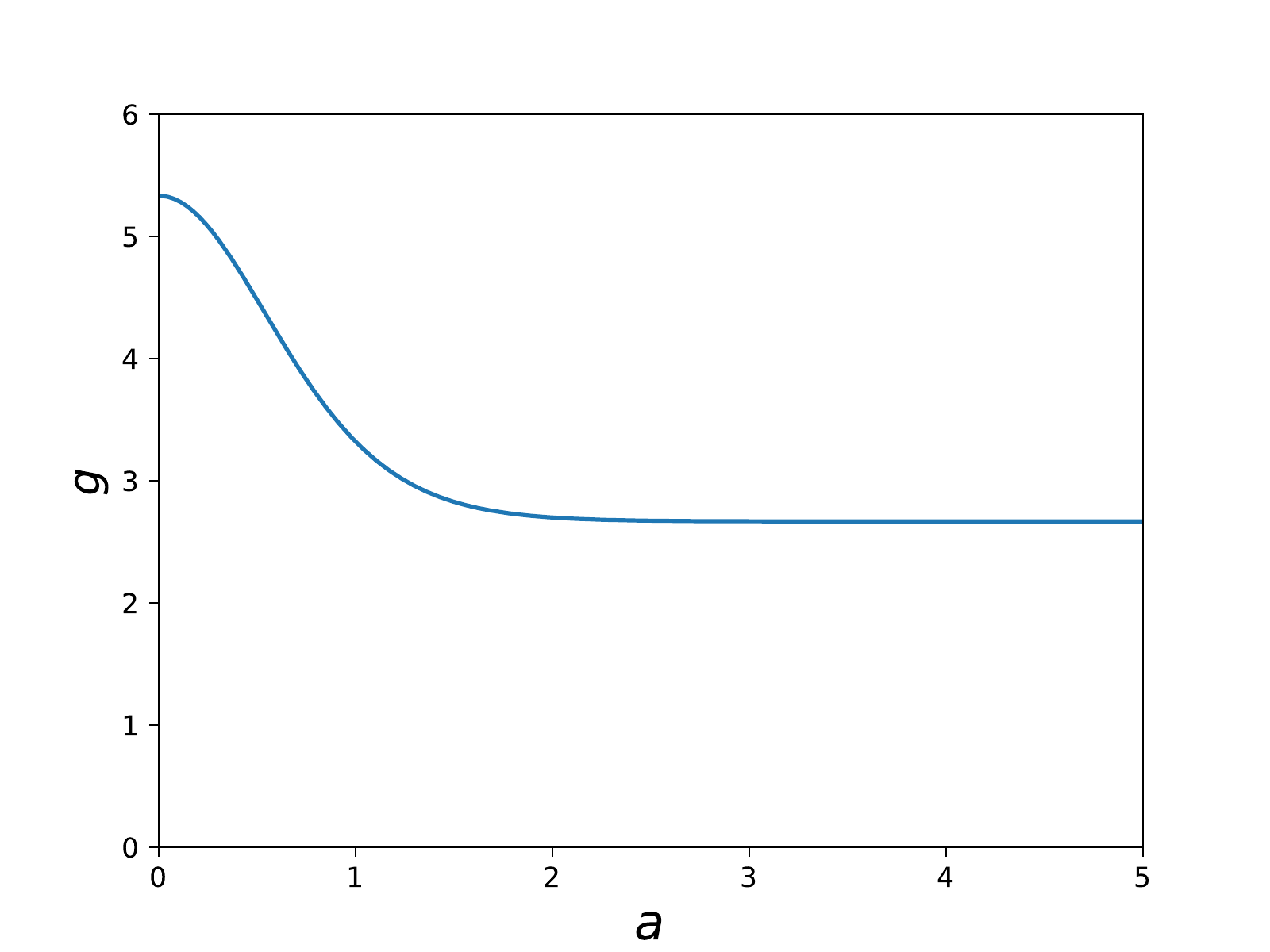}
\includegraphics[scale=0.45]{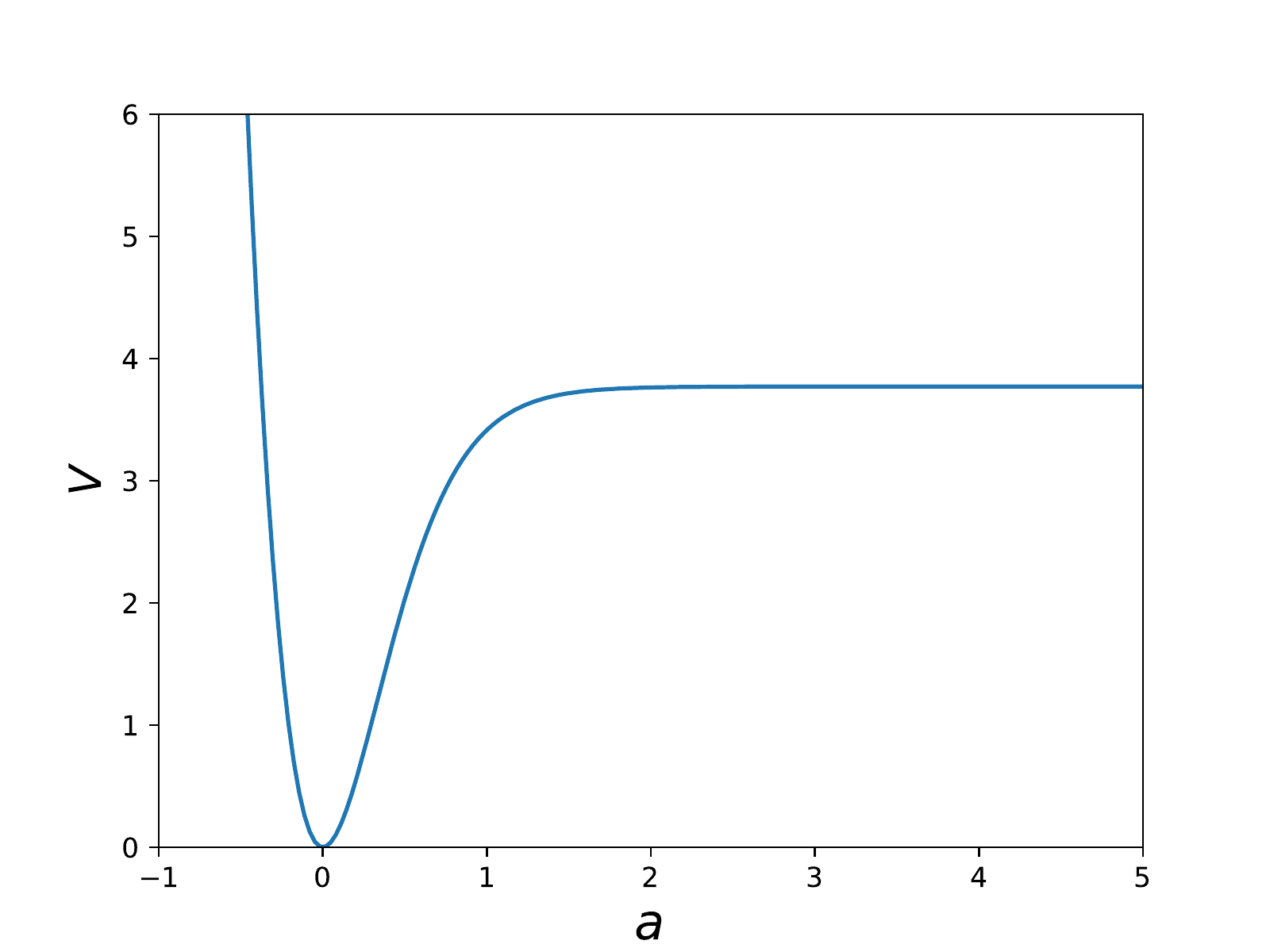}
\caption{The behavior of the functions $g(a)$ and $V(a)$ with $\sigma = \eta = 1$.}
\label{fig: kak gv}
\end{figure}

\begin{figure}[h!]
\centering
\includegraphics[scale=0.4]{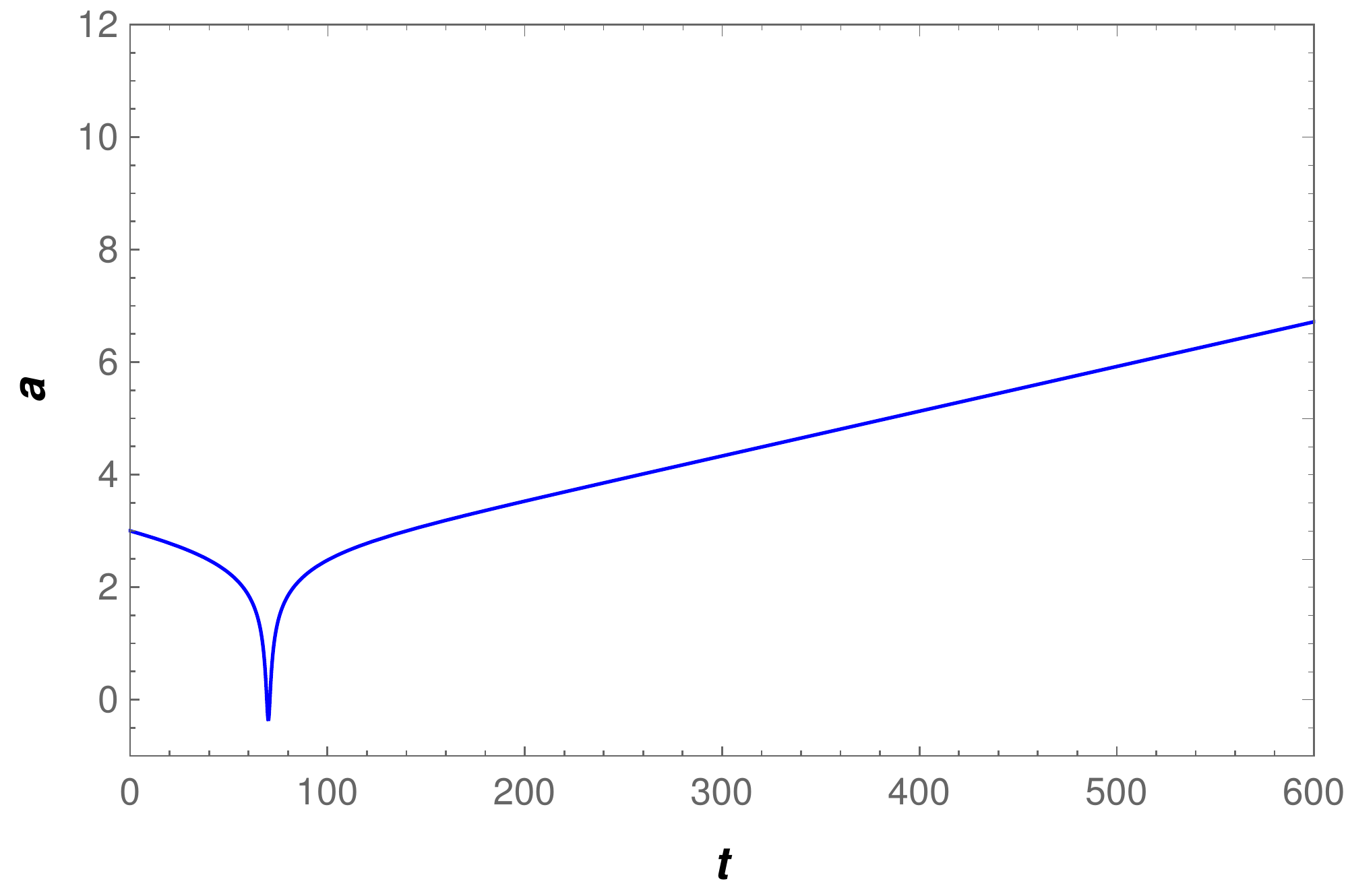}
\includegraphics[scale=0.4]{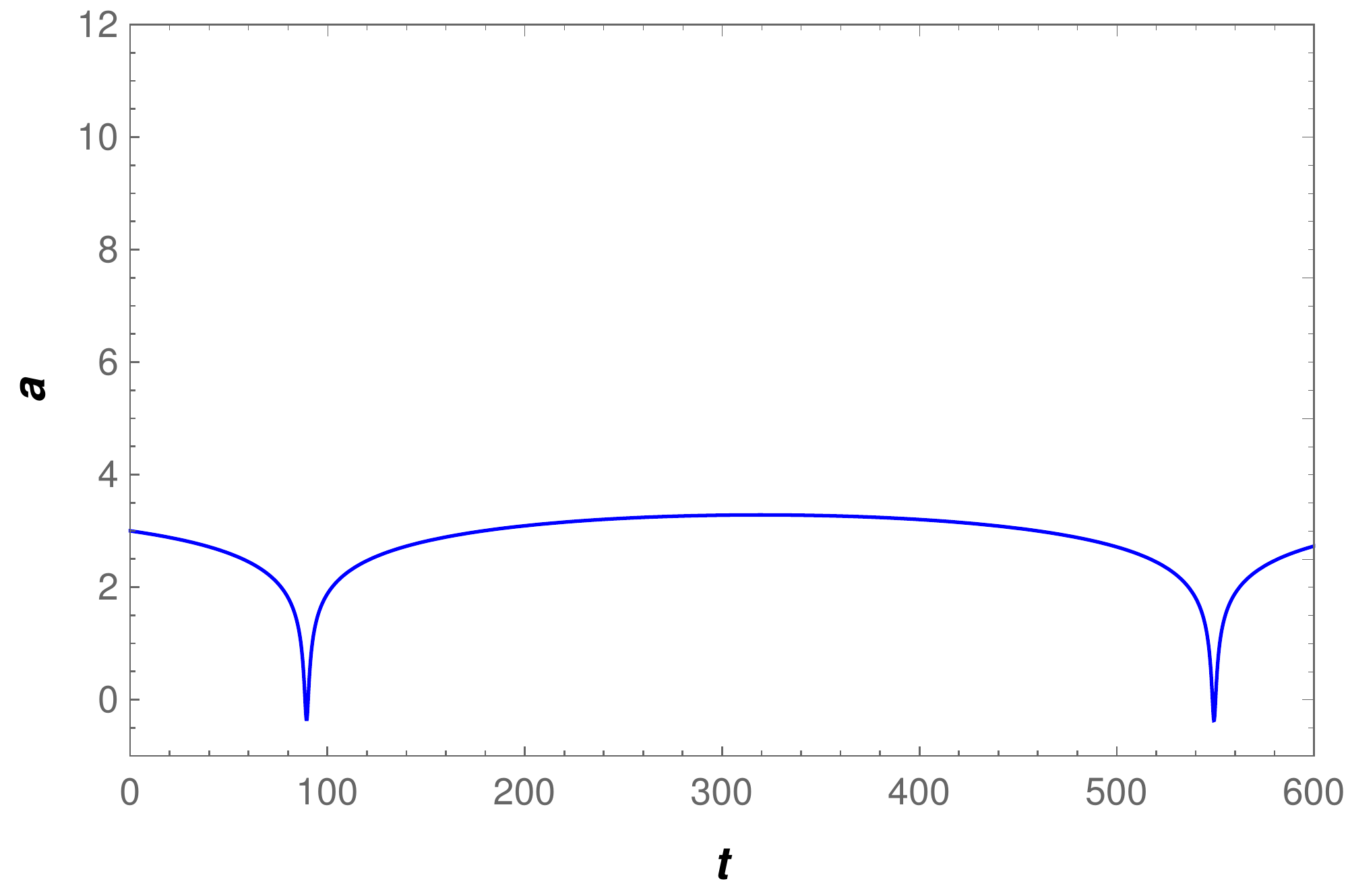}
\caption{The dynamics of $a(t)$ defines the motion of the solitons in the scattering process. In both cases presented here, $\sigma = \eta = 1$. The figure on the left shows an elastic scattering where the solitons started at $a(0) = 3$ with initial velocity $\dot{a}(0)=0.01$. The figure on the right shows a bounded motion obtained for initial velocity $\dot{a}(0) = 0.005$.}
\label{fig: kak belova motion}
\end{figure}

Different initial conditions for $a(t)$ and $\dot{a}(t)$ will imply different types of predicted motion for the solitons: both bound and unbound motion are presented in figure \ref{fig: kak belova motion}.

This approximation is well know, so as its problems. We notice in particular the inconvenient fact that the collective coordinate $a(t)$ assumes negative values in the collision process, which is something unwanted once this coordinate stands for the relative distance between \textit{kink} and \textit{anti}-\textit{kink}. 

In particular, the dynamics of the solitonic pair described by $a(t)$ does not present, in any circumstance, the formation of the resonances observed for certain given initial relative velocities\cite{belova,Anninos:1991un,moshir}. Such a limitation indicates that the use of this collective coordinate only is not enough. 

The existence of a force between \textit{kink} and \textit{anti-kink}, even when these objects are far apart, lead us to consider the excitation of the internal mode of vibration (\ref{eq: mode}) in the dynamics through a new collective coordinate $\xi(t)$.  This coordinate will assume the role of the time dependent amplitude of oscillation of the first excited mode of the (\textit{anti-})\textit{kink} $\epsilon(t,x)=\xi(t)\chi(x)$, with $\chi(x)=\frac{1}{2}\sech{(\sigma x)\tanh{(\sigma x)}}$. We thus consider the possibility of the energy transfer between the translation and vibration modes through the nonlinearities of the model.

Thus, considering\cite{Sugiyama:1979mi} 
\begin{equation}
\label{eq: kak field config}
\phi = \phi_K(x+a(t)) + \phi_{\bar{K}}(x-a(t))-\eta + \xi(t)\left(\chi(x+a(t))-\chi(x-a(t))\right)
\end{equation}
with $\chi(x) = \frac{1}{2}\sech{(\sigma x)}\tanh{(\sigma x)}$, we can proceed as before and integrate the lagrangian density over the spatial coordinates thus obtaining a description in terms of the dynamics of the coordinates $\xi(t)$ and $a(t)$, taken\footnote{We have also considered higher orders in $\xi$ in the lagrangian and no different result than those presented here was observed.} up to quadratic order in $\xi$. Even though this integration is far from being simple to be performed we have successfully obtained, as explained in the appendix \ref{sec: appendix integrals}, the analytical expressions needed for the collective coordinates lagrangian\footnote{The parameters are fixed as $\sigma = \eta = 1$.}
\begin{eqnarray}
\label{eq: lag full}
L&=& \frac{1}{2}\left(M_{0}+I\left(a\right)+{\xi}^2K\left(a\right)+{\xi}J\left(a\right)\right){\dot{a}}^2  + \frac{1}{2}\left( 2+ Q\left(a\right)\right){\dot{\xi}}^2\nonumber\\
&+&{\left(C\left(a\right)+{\xi}N\left(a\right)\right)\dot{\xi}}{\dot{a}}-\left(V\left(a\right)-{\xi}F\left(a\right)+{\xi}^2W\left(a\right)\right),
\end{eqnarray} 
where the coefficients multiplying the canonical variables were found to be
\begin{eqnarray*}
M_{0}&=&\frac{8}{3}\\
I\left(a\right)&=&16a\coth\left(2a\right)\csch^2\left(2a\right)-8\csch^2\left(2a\right)\\
V\left(a\right)&=&8\left[-\frac{2}{3}+2a+\frac{3}{\tanh\left(2a\right)}-\frac{2\left(1+3a\right)}{\tanh^2\left(2a\right)}+\frac{4a}{\tanh^3\left(2a\right)}\right]\nonumber\\
\\
Q\left(a\right)&=&12a\csch\left(2a\right)+24a\csch^3\left(2a\right)-12\coth\left(2a\right)\csch\left(2a\right)\\
C\left(a\right)&=&\pi\sqrt{\frac{3}{2}}\tanh\left(a\right)\sech^2\left(a\right)\\
F\left(a\right)&=&-6\pi\sqrt{\frac{3}{2}}\tanh^2\left(a\right)\left[-1+\tanh\left(a\right)\right]^2\\
J\left(a\right)&=&\sqrt{\frac{3}{2}}\pi\left[1+2\sech^4\left(a\right)-\sech^4\left(a\right)\cosh\left(2a\right)\right]\\
N\left(a\right)&=&\frac{3}{2}\left[15\csch^3\left(2a\right)+9\csch\left(2a\right)\coth^2\left(2a\right)+3\csch\left(2a\right)\right]\\
&+&\frac{3}{2}\left[-6a\csch\left(2a\right)\coth\left(2a\right)-46a\coth\left(2a\right)\csch^3\left(2a\right)\right]\\
&+&\frac{3}{2}\left[-2a\coth^3\left(2a\right)\csch\left(2a\right)\right]\\
K\left(a\right)&=&\frac{3}{2}\left[\frac{28}{15}+160a\csch^3\left(2a\right)+8a\csch\left(2a\right)+192a\csch^5\left(2a\right)\right]\\
&+&\frac{3}{2}\left[-96\coth\left(2a\right)\csch^3\left(2a\right)-16\coth\left(2a\right)\csch\left(2a\right)\right].
\end{eqnarray*}

From this lagrangian, the dynamical equations for the collective coordinates $a(t)$ and $\xi(t)$ are

\begin{eqnarray}
&&\left[C(a)+\xi{N(a)}\right]\ddot{a} +\left[2+Q(a)\right]\ddot{\xi} + Q'(a)\dot{a}\dot{\xi}
+\left[C'(a)-\frac{1}{2}J(a)+\xi\left(N'(a)-K(a)\right)\right]\dot{a}^2\nonumber\\
&+&2\xi{W(a)}-F(a)=0 \\
&& \left[M_{0}+I(a)+\xi{J(a)+{\xi}^2K(a)}\right]\ddot{a}+\left[C(a)+\xi{N(a)}\right]\ddot{\xi}+\left[N(a)-\frac{1}{2}Q'(a)\right]{\dot{\xi}}^2\nonumber\\
&+&\frac{1}{2}\left[I'(a)+{\xi}J'(a)+{\xi}^2K'(a)\right]{\dot{a}}^2 +\left[J(a)+2{\xi}K(a)\right]{\dot{a}}{\dot{\xi}}+V'(a)+{\xi}^2W'(a)-{\xi}F'(a)=0.
\end{eqnarray}

These equations can be written as a system of the form\cite{belova} $M \mathbf{a} = \mathbf{f}$, where $\mathbf{a} = (\ddot{a},\ddot{\xi})$, with
\begin{equation}
M = \left(
\begin{array}{cc}
\left[C(a)+\xi{N(a)}\right]&\left[2+Q(a)\right]\\
\left[M_{0}+I(a)+\xi{J(a)+{\xi}^2K(a)}\right]&\left[C(a)+\xi{N(a)}\right]
\end{array}
\right)
\end{equation}
and $\mathbf{f}$ stands for all the other quantities in terms of the collective coordinates and their derivatives. For nontrivial solutions of this system it is required that
\[
\left(M_{0}+I\left(a\right)+{\xi}^2K\left(a\right)+{\xi}J\left(a\right)\right)\left( 2+ Q\left(a\right)\right) - \left(C+\xi N\right)^2\neq 0.
\]

An analysis of this condition shows that the requirement $C=N=0$ is needed for its validity for any values of $a(t)$ and $\xi(t)$. 

In the recent literature on this subject, an important contribution\cite{Takyi:2016tnc} was given concerning the correction of the term here labeled as $F(a)$, which was mistakenly written in \cite{Sugiyama:1979mi}. In the appendix \ref{sec: appendix integrals} we discuss how the integrals leading to this term can be done using the method of residues and this approach confirms the result presented in \cite{Takyi:2016tnc}. Although of remarkable importance, this correction was not sufficient to produce better results within the approximation considered in \cite{Takyi:2016tnc}, and in general throughout the literature, which considers, in the lagrangian (\ref{eq: lag full}) given above,  $Q = K = J = 0$ and $W = 3$, besides the already mentioned necessity of taking $C = N = 0$. 

While the choice of $W = 3$ can be justified by the fact that it really makes the approximation worst if otherwise chosen and thus it is set as the constant value it assumes in the limit $a\to \infty$, the reason behind the vanishing of the functions $Q$, $K$ and $J$ given is generally not very clearly justified. 

\begin{figure}[h!]
  \centering
  \begin{subfigure}{0.49\textwidth} 
    \includegraphics[width=\textwidth]{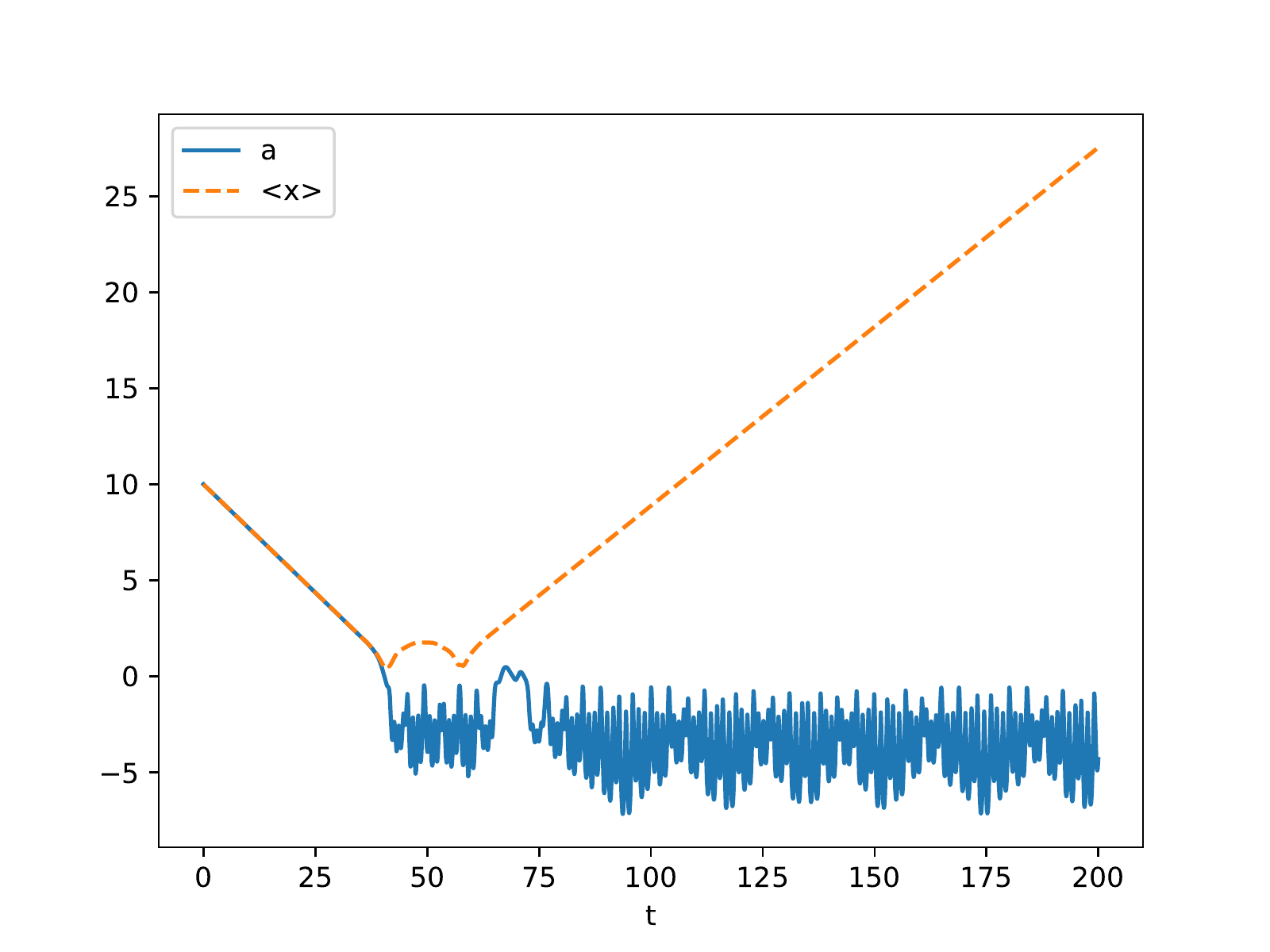}
    \caption{A resonance is observed for the initial velocity of $0.2249$.}
    \label{fig: v02246_takyi}
  \end{subfigure}
  \vspace{1em} 
  \begin{subfigure}{0.49\textwidth} 
    \includegraphics[width=\textwidth]{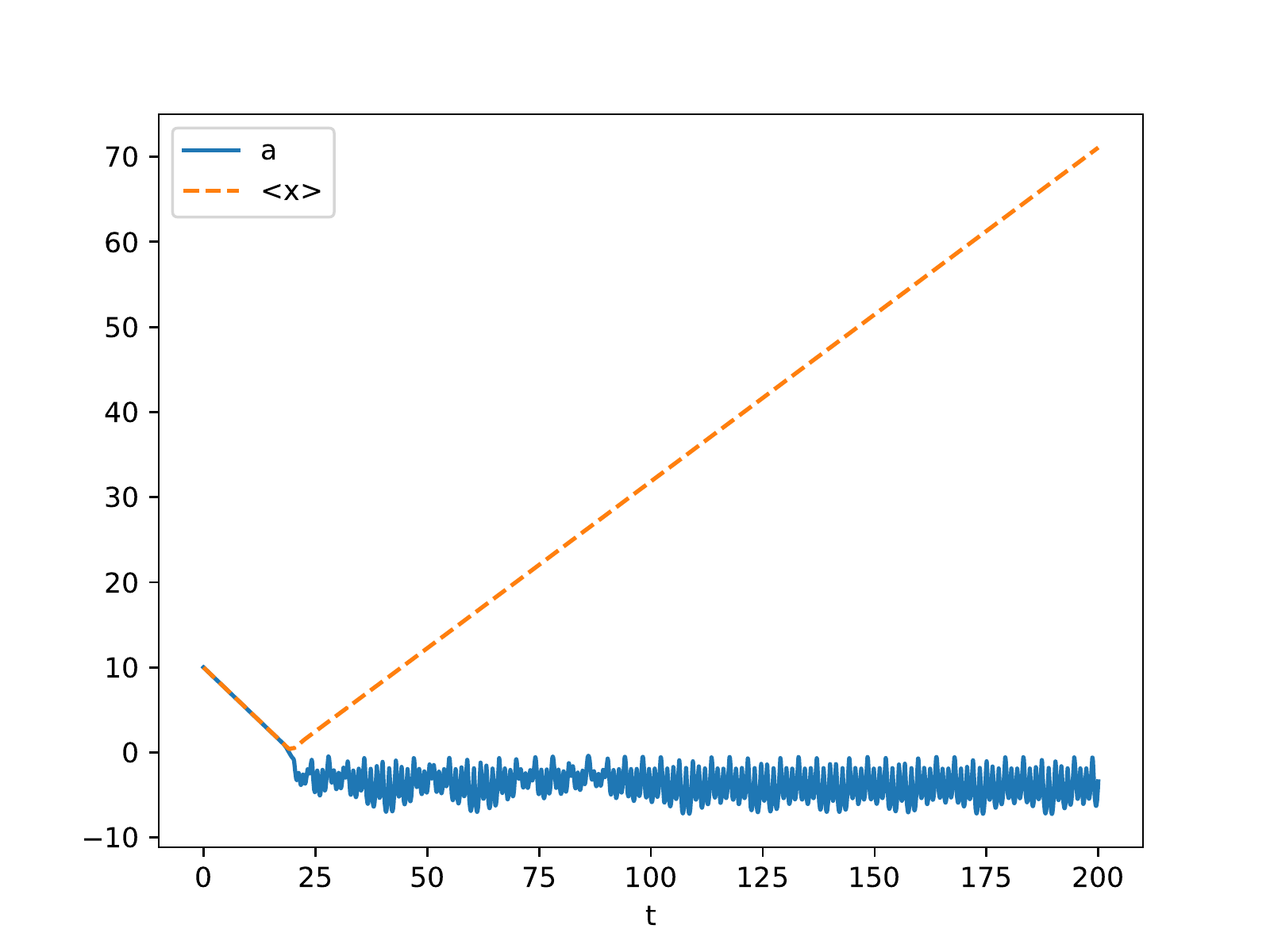}
    \caption{With velocity of $0.5$ the solitons undergo an elastic collision.}
    \label{fig: v05_takyi}
  \end{subfigure}
  \caption{The full line gives the dynamics of the coordinate $a(t)$ within the approximation where $Q = K = J = 0$ and the dashed line shows the dynamics of the system obtained from the full numerical simulation.}
  \label{fig:takyi}
\end{figure}

As discussed in \cite{Takyi:2016tnc}, even with the correction of the term labeled as $F$, the approximation considered there and throughout the literature - as far as we know - not only leads to a physically non-acceptable negative value of $a(t)$ when the solitons collide but also presents some problems in predicting elastic collisions when they are expected in a situation where high initial relative velocities are considered. These results are reproduced in figure \ref{fig:takyi} where for each case we have also given the center of mass of the system $\langle x \rangle$ calculated from the full numerical simulation as
\begin{equation}
\langle x \rangle = \frac{\int dx\; x\mathcal{E}}{\int dx\; \mathcal{E}}
\end{equation}
where $\mathcal{E}$ stands for the energy density of the field configuration (\ref{eq: kak field config}):
\begin{equation}
\mathcal{E} = \frac{1}{2}\left(\frac{\partial \phi}{\partial t}\right)^2+\frac{1}{2}\left(\frac{\partial \phi}{\partial x}\right)^2 + \mathcal{U}.
\end{equation}

\subsection{The inclusion of more terms in the approximation}

Here we are inclined to state that in fact, the problems which appear in the construction of the collective coordinates approximation presented in \cite{Takyi:2016tnc} and reproduced in figure \ref{fig:takyi}, no longer exist when most of these terms are taken into account, i.e., when $Q$, $K$ and $J$ are included. Thus we consider the space of the parameters $(a,\xi)$ to be defined by a diagonal metric so that the lagrangian reads
\begin{equation}
\label{eq: lag final}
L= \frac{1}{2}\left[M_{0}+I\left(a\right)+{\xi}^2K\left(a\right)+{\xi}J\left(a\right)\right]{\dot{a}}^2  + \frac{1}{2}\left[ 2+ Q\left(a\right)\right]{\dot{\xi}}^2-\left[V\left(a\right)-{\xi}F\left(a\right)+3\,{\xi}^2\right].
\end{equation}   

\begin{figure}[h!]
  \centering
  \begin{subfigure}{0.49\textwidth} 
    \includegraphics[width=\textwidth]{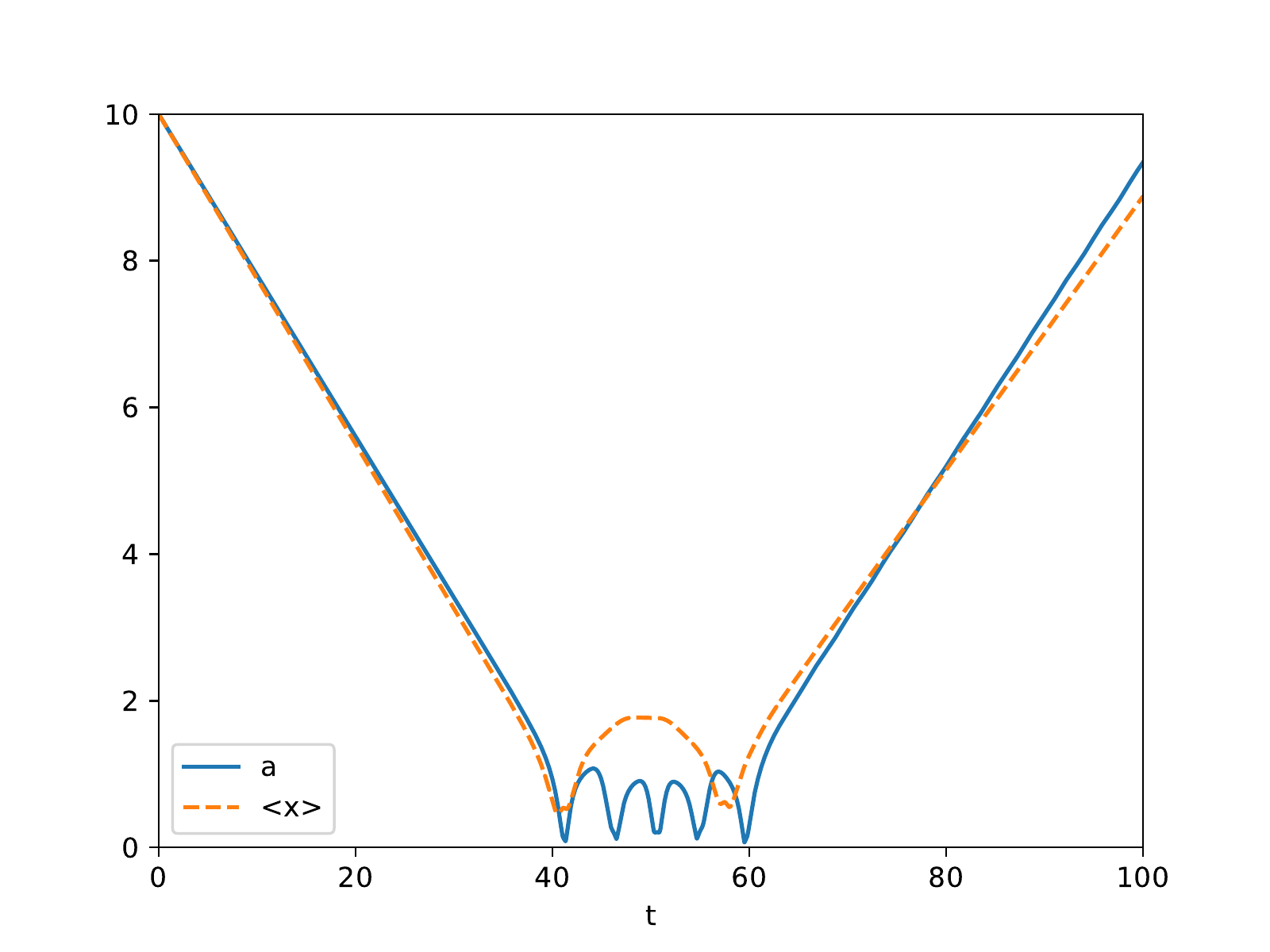}
    \caption{The solution for $a(t)$ with the initial velocity of $0.2249$.}
    \label{fig: v02249_takyi}
  \end{subfigure}
  \vspace{1em} 
  \begin{subfigure}{0.49\textwidth} 
    \includegraphics[width=\textwidth]{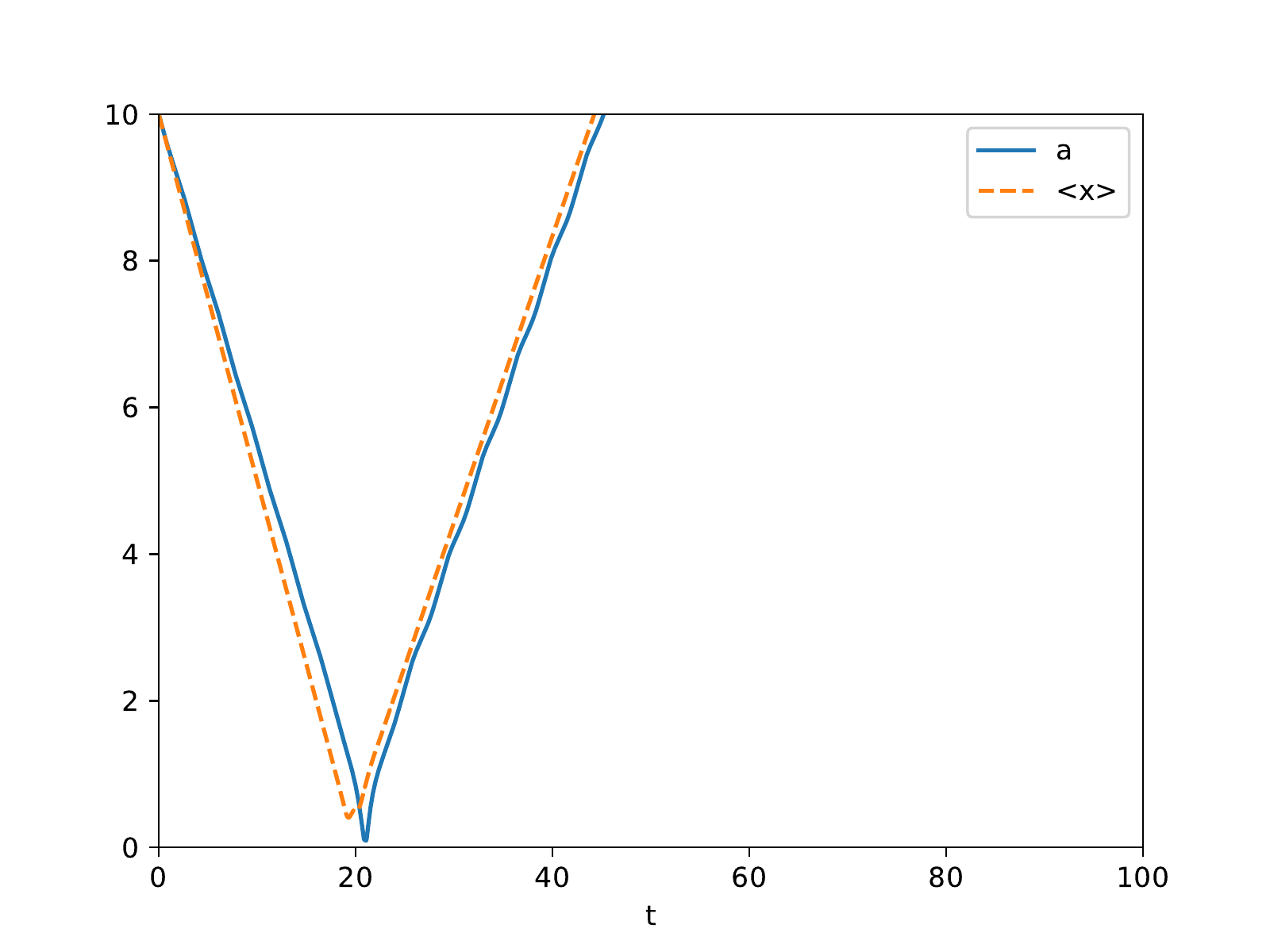}
    \caption{The solution for $a(t)$ with initial velocity of $0.5$}
    \label{fig: v05_takyi}
  \end{subfigure}
  \caption{The collective coordinate approximation seems to reproduce some of the fundamental properties of the dynamics of the \textit{kink}/\textit{anti}-\textit{kink} scattering. The resonance window for velocity $0.2249$ can be seen in this approximation.}
  \label{fig: coll coord a}
\end{figure}

In figure \ref{fig: coll coord a} below we show the numerically obtained solution of the Euler-Lagrange equations for $a(t)$ for some choices of initial relative velocity. 

For these velocities, we may say that the collective coordinates approximation exhibits some good agreement with what is observed in the full numerical simulation, described by $\langle x \rangle$. Also, we can see that the resonance phenomenon is at a certain degree, described by the approximation. This may indicates that indeed, the two degrees of freedom, namely, the translation and vibration modes, play the crucial role - as expected - in the scattering process.

One important remark to be made is that the coefficient labeled as $Q(a)$ in the lagrangian (also found explicitly in appendix \ref{sec: appendix integrals}) is different than the one labeled the same in \cite{Sugiyama:1979mi}. While the latter is divergent for $a\rightarrow 0$, i.e., when the solitons collide, the function $Q$ found here has value $Q(a\to 0) = -2$, which makes the kinetic part associated to the vibrations in the lagrangian (\ref{eq: lag final}) vanishes. This is certainly of great importance for the results just presented.

\begin{figure}[h!]
  \centering
  \begin{subfigure}{0.49\textwidth} 
    \includegraphics[width=\textwidth]{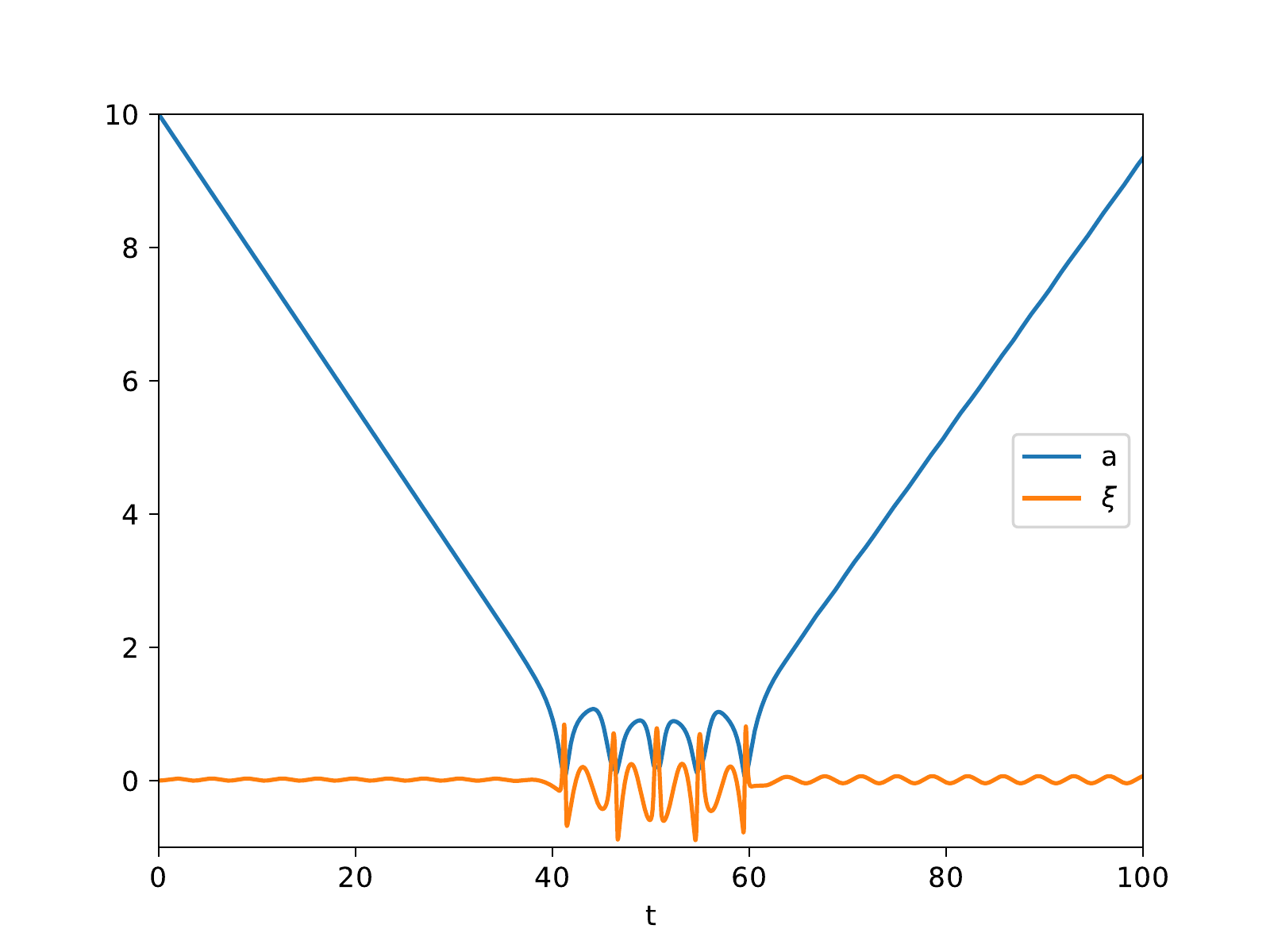}
    \caption{The solutions for $a(t)$ and $\xi(t)$ for the initial velocity of $0.2249$.}
    \label{fig: v02249_a_xi}
  \end{subfigure}
  \vspace{1em} 
  \begin{subfigure}{0.49\textwidth} 
    \includegraphics[width=\textwidth]{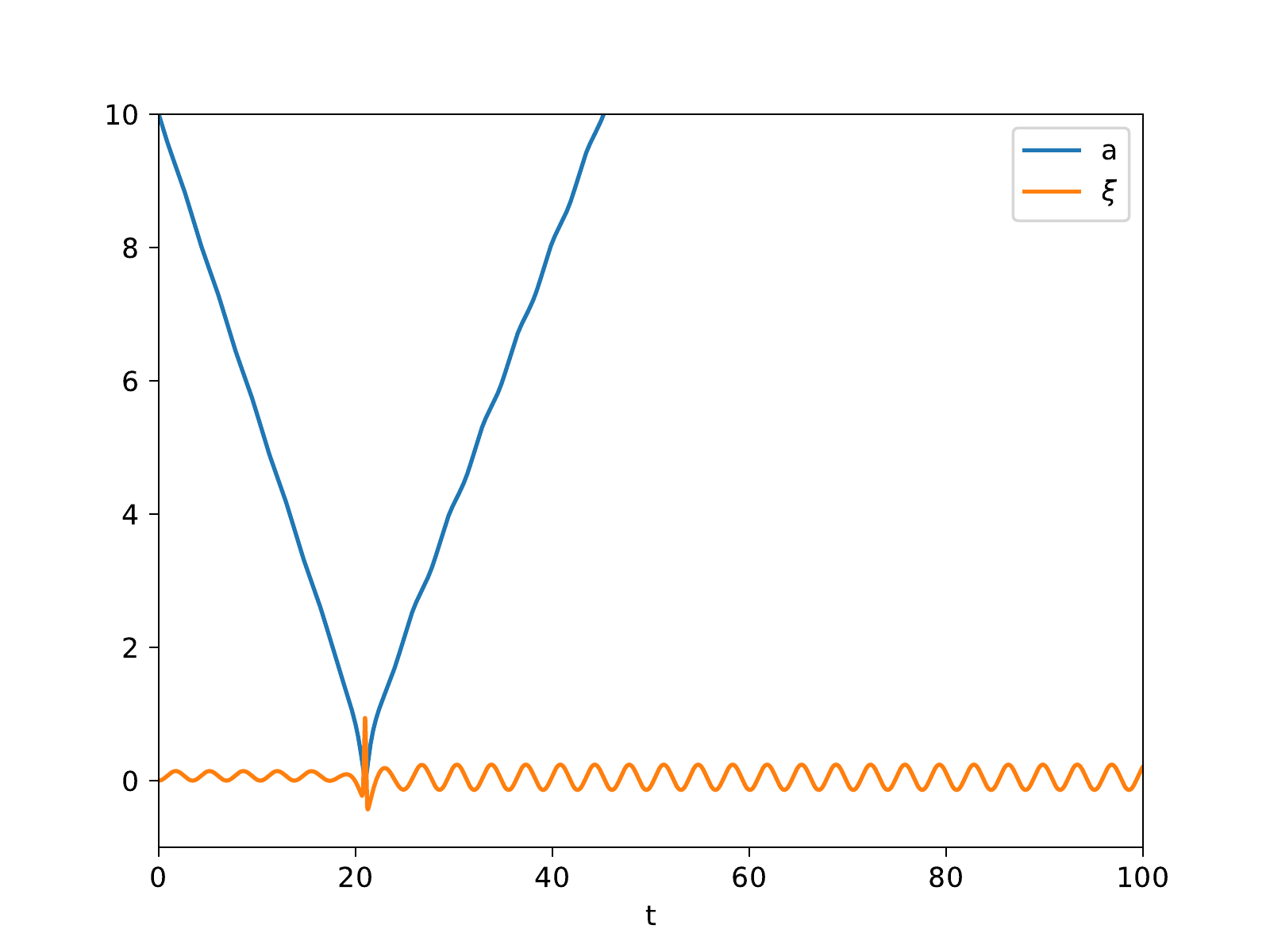}
    \caption{The solution for $a(t)$ with initial velocity of $0.5$}
    \label{fig: v05_a_xi}
  \end{subfigure}
  \caption{It is interesting to note that the amplitude of the oscillation $\xi(t)$ increases after the scattering.}
  \label{fig: coll coord a xi}
\end{figure}

In order to get a more detailed picture of the role of the translation and vibration modes, in figure \ref{fig: coll coord a xi} we show the solution of $\xi(t)$ together with $a(t)$.
 
 We notice that the amplitude of the oscillation $\xi(t)$ is larger after the scattering. This seems to indicate, together with the previously discussed fact that no considerable amount of energy is lost in the scattering process, that part of the initial energy from the translational motion was transfered to the vibrational mode. In fact, this can be also seen in figures \ref{fig: kak bounce v04} and \ref{fig: kak field zero v04}, from the full numerical simulation results.

The energy transfer process between the two degrees of freedom can be seen by defining the total energy of the collective coordinate system as the sum of two functions, $E = E_a + E_\xi$, where
\begin{eqnarray}
E_a &=& \frac{1}{2}\left(M_{0}+I\left(a\right)\right){\dot{a}}^2  + V\left(a\right)\\
E_\xi &=& \frac{1}{2}\left( 2+ Q\left(a\right)\right){\dot{\xi}}^2+\left({\xi}^2K\left(a\right)+{\xi}J\left(a\right)\right){\dot{a}}^2 -{\xi}F\left(a\right)+3\,{\xi}^2.
\end{eqnarray}

\begin{figure}[h!]
  \centering
  \begin{subfigure}{0.49\textwidth} 
    \includegraphics[width=\textwidth]{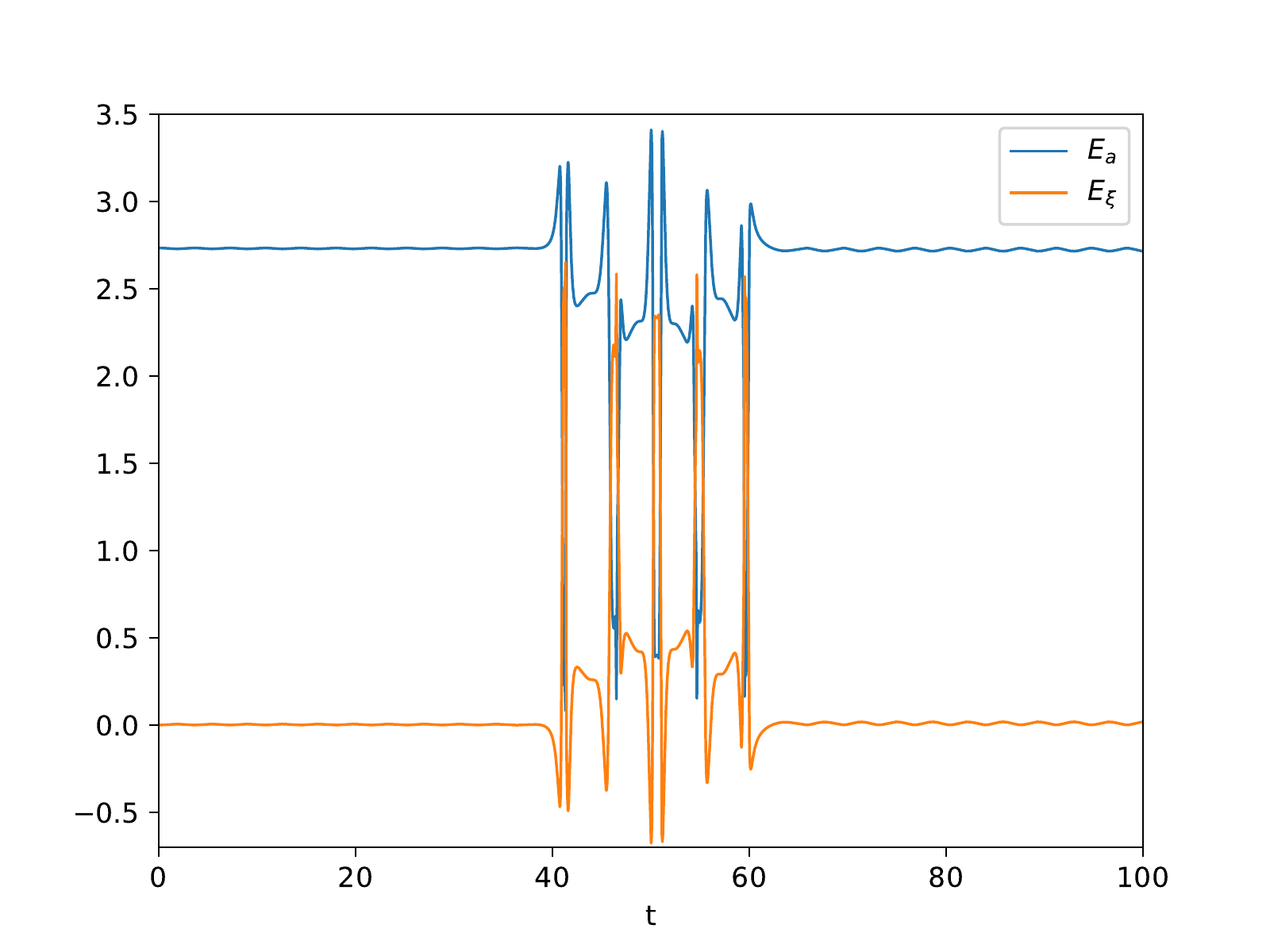}
    \caption{The energy of the collective coordinates system for velocity equal to $0.2249$.}
    \label{fig: v02249_engy}
  \end{subfigure}
  \vspace{1em} 
  \begin{subfigure}{0.49\textwidth} 
    \includegraphics[width=\textwidth]{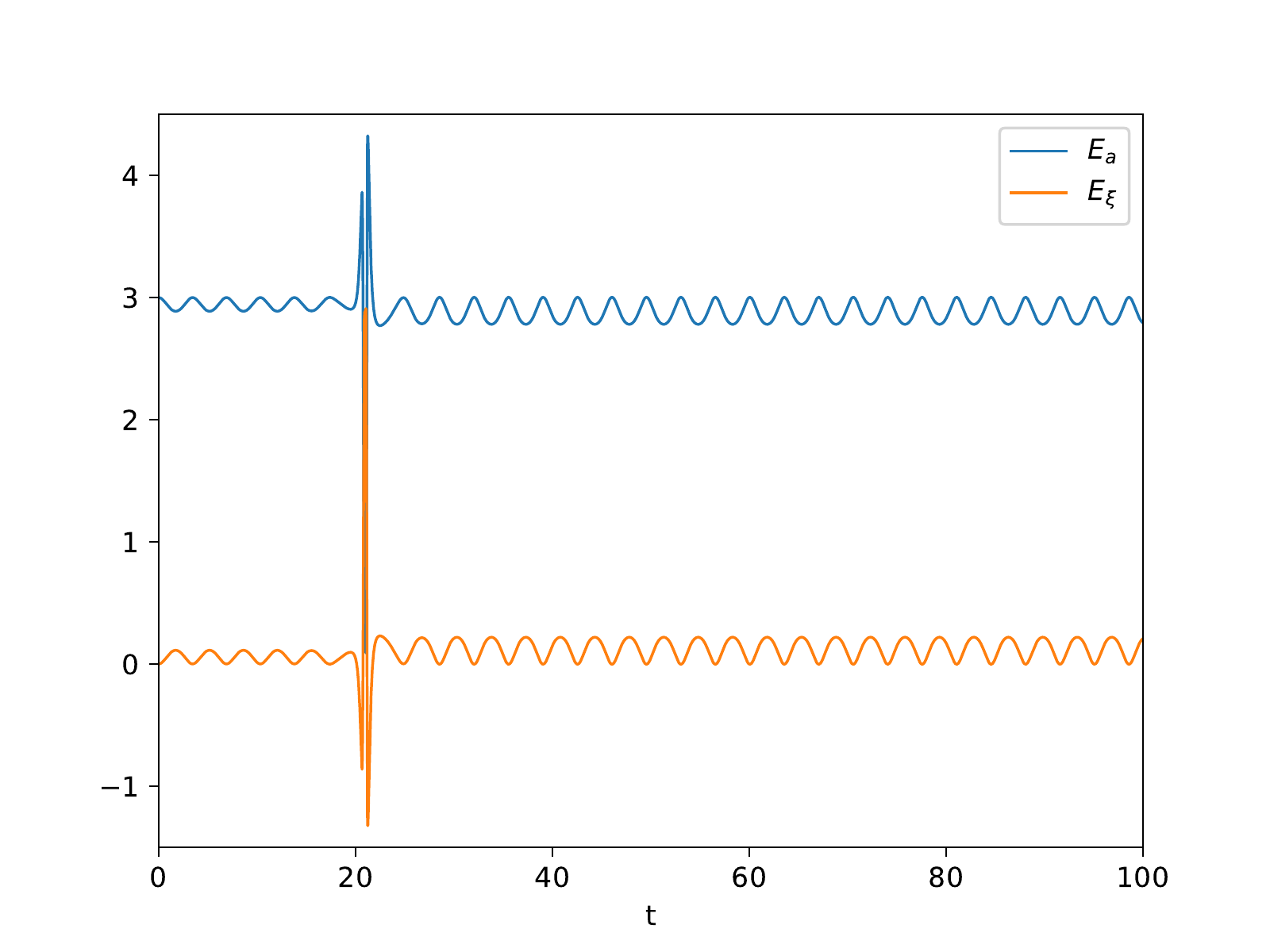}
    \caption{The energy of the collective coordinate system for velocity equal to $0.5$}
    \label{fig: v05_engy}
  \end{subfigure}
  \caption{The translational and vibrational modes become evident with respect to the exchange of energy if it is separated in terms of $E_a$ and $E_\xi$.}
  \label{fig: coll coord engy}
\end{figure}

In figure \ref{fig: coll coord engy} we show the behaviour of these functions for two cases, with velocities $v=0.2249$ and $v=0.5$.  A qualitative analysis shows that when the solitons are getting closer, part of the energy stored in the internal mode will be transfered to the translational motion and the solitons will accelerate towards each other. Next, the energy is then transfered to the internal mode and the relative velocity gets lower. This can be lower than the ``escape velocity'', i.e., the minimum velocity the solitons need to perform an elastic collision and consequently the solitons get trapped. In another situation, if the initial velocity is enough, the energy can flow back to the transational mode so that, at some stage, the solitons will recover sufficient velocity to escape; that is the case of the resonance phenomenon.

\begin{figure}[h!]
  \centering
  \begin{subfigure}{0.49\textwidth} 
    \includegraphics[width=\textwidth]{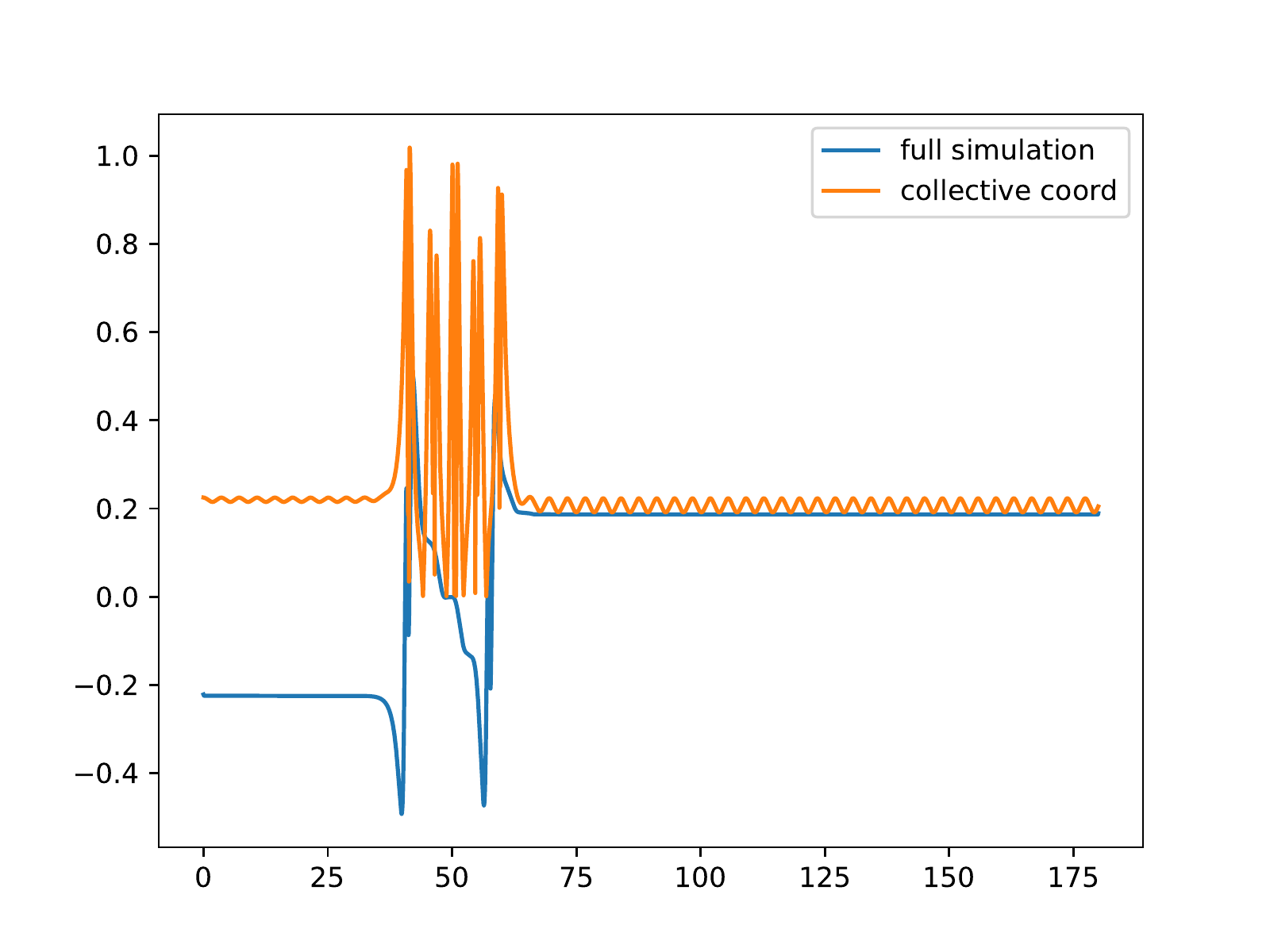}
    \caption{The velocity profile of the soliton during the scattering process with initial velocity equal to $0.2249$.}
    \label{fig: v02249_engy}
  \end{subfigure}
  \vspace{1em} 
  \begin{subfigure}{0.49\textwidth} 
    \includegraphics[width=\textwidth]{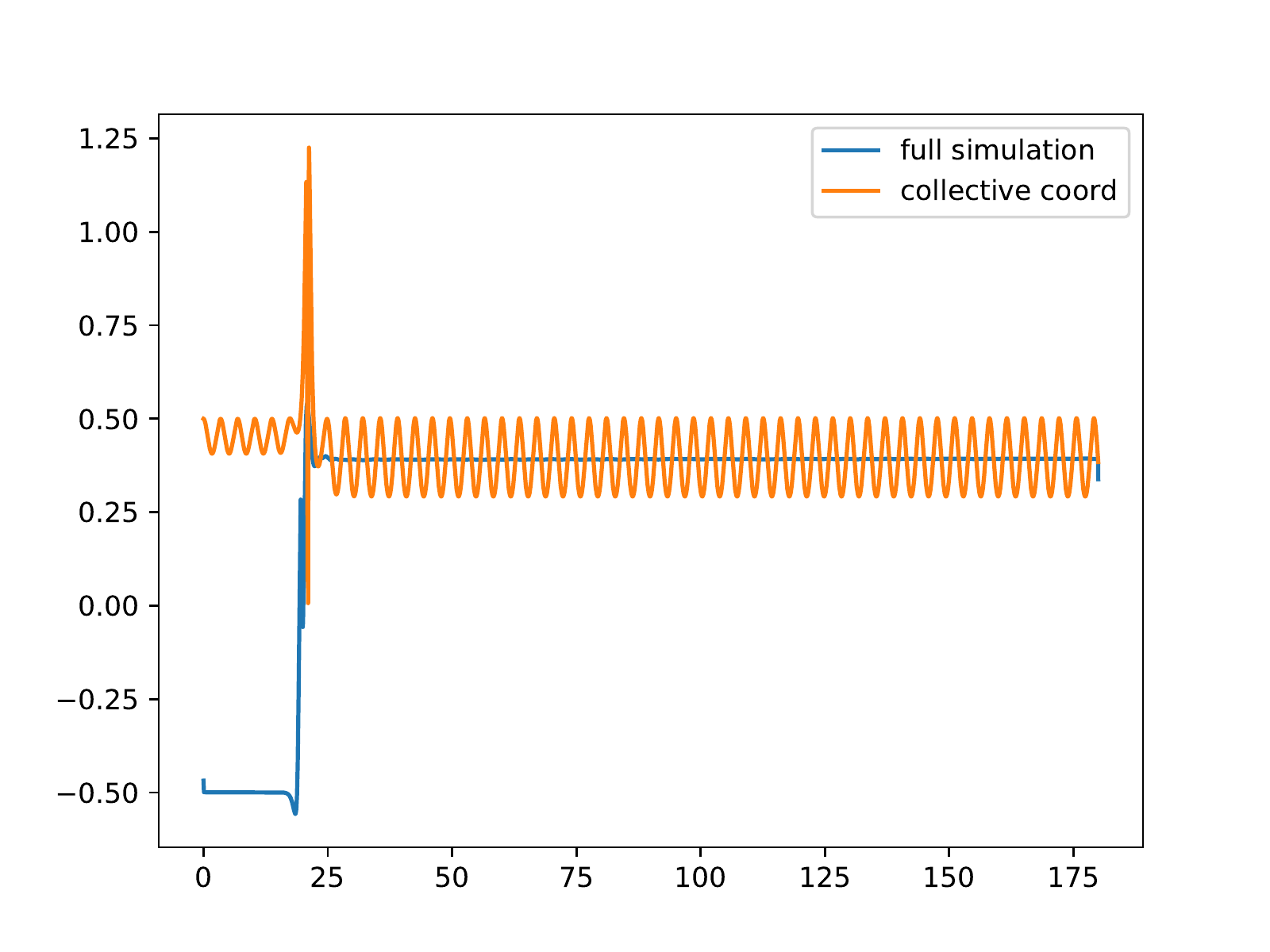}
    \caption{The velocity profile of the soliton during the scattering process with initial velocity equal to $0.5$}
    \label{fig: v05_engy}
  \end{subfigure}
  \caption{The velocity profile of the solitons given from the full numerical simulation and using the collective coordinates, in which case we present the absolute value.}
  \label{fig: avg vel}
\end{figure}

 In figure \ref{fig: avg vel} we present the behaviour of the velocity of the solitons during the scattering processes, calculated using the full numerical simulation results and also the one obtained from the collective coordinates approximation (which is given in terms of its absolute value). We notice that indeed, by looking at the full simulation results for the cases shown and for all other we have seen, the solitons leave the collision with lower velocity than that with which they entered.

As seen from figure \ref{fig: avg vel}, the collective coordinates approximation captures the fact that the velocity gets lower after the collision by increasing the solitons back and fourth oscillation.

\section{Conclusions} \label{3}

We have shown that through the technique presented in the appendix \ref{sec: appendix integrals} all the coefficients which appear in the lagrangian of the collective coordinates system obtained with (\ref{eq: kak field config}) for the scattering of \textit{kink} and \textit{anti}-\textit{kink} of the $\phi^4$ model can be found analytically. These results enabled us to push the approximation method to ``its best possible performance'', so that no further inclusion of terms aiming to correct the approximation, as those described in \cite{Takyi:2016tnc}, are needed. The only assumption made are that $W$ is taken to have the constant value of $W=3$ and $N=C=0$, which are already known in the literature and well justified.

The collective coordinates approximation as described here allows for the appearance of the resonance windows which are seen in the full numerical simulation. Moreover, also elastic scattering can be described through this approximation. Nevertheless, it must be remarked that the effectiveness of the approximation is highly dependent on the value of the initial relative velocity between the solitons. We have seen in our numerical experiments that for certain values of this parameter, what is observed for $a(t)$ can be very different from what is found for $\langle x \rangle$. 

The relative success of the collective coordinates approximation involving a translational and a vibrational mode as presented here seems to indicate that indeed it is the interchange of energy between these modes that results in the resonances found in the scattering of $\phi^4$ solitons. The collective coordinate system exhibits a recurrence phenomenon where almost all the energy, after the collision, is distributed back to the translation of the solitons which allows for them to undone the temporary bound system, when it is formed. In our numerical experiments we have seen that in all cases considered, some of the energy absorbed in the vibrational mode during the collision remains there, so that the solitons have some wiggling after the collision.


\appendix
\section{Perturbation of the \textit{kink} solution}
\label{sec: app pert}
For completeness we present in this section a detailed calculation concerning the dynamics of the perturbation of a \textit{kink} solution of the $\phi^4$ model.

Writing $\phi_0$ as the (\textit{anti-})\textit{kink} solution we consider a configuration given by

\begin{eqnarray}
\label{eq: pert conf}
\phi\left(x,t\right)= \phi_{0}\left(x\right) + \epsilon\left(x,t\right)
\end{eqnarray} 
where $\vert \epsilon \vert \ll \phi_0$ is the perturbation field. This configuration will satisfy the dynamical equation (\ref{eq: eom phi4}) for linear order in the perturbation if  

\begin{equation}
\label{eq: epsilon eq}
-\ddot{\epsilon} + {\epsilon}'' -\lambda\epsilon{\eta}^2\left[3\tanh^2\left(\sigma{x}\right)-1\right]=0 
\end{equation}
holds. 

This equation has solutions of the form $\epsilon\left(x,t\right)= \chi\left(x\right)e^{\imath{\omega}t}$ given that $\chi$ satisfies the eigenvalue equation



\begin{equation}
\label{eq: eigen equation}
\frac{d^2\chi}{dx^2} + \left[E+\frac{U_{0}}{\cosh^2\left(\sigma{x}\right)}\right]\chi\left(x\right)=0,
\end{equation}
with $E={\omega}^2-4{\sigma}^2$ and $U_{0}= 6{\sigma}^2$.

With the change of variable $y=\frac{1}{2}\left[1-\tanh\left(\sigma{x}\right)\right]$ and defining $\frac{E}{{\sigma}^2}=-{\varepsilon}^2$ e $\frac{U_{0}}{{\sigma}^2}=s\left(s+1\right)$, this equation becomes


\begin{eqnarray}\label{flu9}
y\left(1-y\right)\frac{d^2\chi}{dy^2} + \left(1-2y\right)\frac{d\chi}{dy} +\left[s\left(s+1\right)-\frac{{\varepsilon}^2}{4y\left(1-y\right)}\right]\chi=0.
\end{eqnarray}

In order to obtain a well-behaved solution we take into account the fact that for $x\to \infty$ we get $y \to 0$ and this equation reads, at this limit\footnote{ We have that $
x \to \infty     ,   y \to 0   \qquad
x \to -{\infty}  ,   y \to 1   \qquad
x \to  {0}       ,   y \to  \frac{1}{2}.$}

\begin{eqnarray}\label{flu10}
y^2{\chi}''+y{\chi}' -\frac{{\varepsilon}^2}{4}\chi =0,
\end{eqnarray} 

and thus with ${\chi}\;{\sim}\;{y}^{\alpha}$, the asymptotic solution is

\begin{eqnarray}\label{flu11}
\chi = C_{1}{y}^{\frac{\varepsilon}{2}}+ C_{2}{y}^{-\frac{\varepsilon}{2}},
\end{eqnarray} 
and $C_{2}$ is taken to be zero.


Finally, taking 
\begin{eqnarray}
\label{eq: change 2}
\chi = \left[y\left(1-y\right)\right]^\frac{\varepsilon}{2}W\left(y\right),
\end{eqnarray} 
we end up with the hypergeometric equation\cite{arf}

\begin{eqnarray}
\label{eq: hyper equation}
y\left(1-y\right)W''+ \left(\varepsilon +1\right)\left(1-2y\right)W' -\left[\left(\varepsilon-s\right)\left(s+\varepsilon+1\right)\right]W=0,
\end{eqnarray} 
whose solution is well known\cite{arf} to be

\begin{eqnarray}
\label{eq: hyper}
\chi\left(y\right)=\left[y\left(1-y\right)\right]^\frac{\varepsilon}{2} {_{2}}F_{1}\left[\varepsilon-s,s+\varepsilon+1,1+\varepsilon,\frac{1}{2}\left(1-y\right)\right].
\end{eqnarray}  

Then, it is required that $\varepsilon-s =-n$, so that the polynomial character of the solution holds. This gives 
\begin{eqnarray}
\label{hyper spectrum}
E_{n}= -{\sigma}^2\left(2-n\right)^2
\end{eqnarray} 
and ${\varepsilon}^2=-\frac{E}{{\sigma}^2}$ implies $\varepsilon= 2-n>0$, and therefore only two values of $n$ are allowed: $n=0,1$.

\subsection{The perturbation field}

We start by rewriting (\ref{eq: hyper}) as

\begin{eqnarray}\label{flu19}
{\chi}_{n}= \left[y\left(1-y\right)\right]^\frac{2-n}{2} \frac{\Gamma\left(3-n\right)\Gamma\left(5\right)}{\Gamma\left(5-n\right)\Gamma\left(3\right)}\left(\frac{1}{2}\left(y-1\right)\right)^{n}.
\end{eqnarray}

Reintroducing the original variables we find 
\begin{eqnarray}\label{flu20}
{\chi}_{0}= \frac{1}{4}\sech^2\left(\sigma{x}\right),
\end{eqnarray}
and the first excited solution reads
\begin{equation}
\epsilon = \frac{1}{4}\sech^2\left(\sigma{x}\right),
\end{equation}
having $E_0 = -4\sigma^2$ and thus a zero frequency of oscillation $\omega$. This is called a translational mode.

For $n=1$ we find
\begin{eqnarray}\label{flu21}
{\chi}_{1}= \frac{1}{2} \sech\left(\sigma{x}\right){\tanh\left(\sigma{x}\right)}
\end{eqnarray}
and with $E_1 = -\sigma^2$ and ${\omega}= \sqrt{3}\sigma$, giving a non-trivial excitation of the \textit{kink}:
\begin{equation}
\epsilon = \frac{1}{2} \sech\left(\sigma{x}\right){\tanh\left(\sigma{x}\right)}\cos\left(\sqrt{3}{\sigma}t\right).
\end{equation}

\section{On the calculation of the integrals appearing in the effective lagrangian}
\label{sec: appendix integrals}
In this section we show the explicit calculation of two particular terms appearing in the construction of the effective lagrangian for the collective coordinates. The method can be used to compute all the integrals involved.

We start by considering the term defined as
\begin{equation}
Q(a) = -3\int_{-\infty}^{+\infty}dx\;\tanh{(x+a)}\sech{(x+a)}\tanh{(x-a)}\sech{(x-a)}.
\end{equation}

With $\omega = x+a$ and $\omega' = x-a = \omega - 2a$ this is rewritten as
\begin{equation}
Q(a) = -3\int_{-\infty}^{+\infty}d\omega\;\tanh{\omega}\sech{\omega}\tanh{(\omega-2a)}\sech{(\omega-2a)}.
\end{equation}

In order to compute this integral we consider it to be part of the integral along a closed curve in the complex plane of the function
\begin{equation}
\label{eq: compl func 1}
f(z) = z \tanh{z}\sech{z}\tanh{(z-2a)}\sech{(z-2a)}
\end{equation}
where $z = \omega +i\phi \in \mathbf{C}$. This complex integral reads
\[
\oint_{\gamma} dz\;f(z) = \int_{-R}^R d\omega\;f(\omega) +i\int_0^{i\pi} d\phi\;f(R + i\phi) - \int_{-R}^R d\omega\;f(\omega + i\pi) - i\int_0^{i\pi} d\phi\;f(-R + i\phi),
\]
and the path $\gamma$ is a rectangle from $-R$ to $R$ in the real axis and from $0$ to $i\pi$ in the imaginary axis.

Next we notice that the integration over the paths with constant $R$ will give no contribution to the result in the limit $R\to \infty$ and therefore
\[
\lim_{R\rightarrow \infty}\oint_\gamma dz\;f(z) = \int_{-\infty}^{+\infty} d\omega\;f(\omega)- \int_{-\infty}^{+\infty} d\omega\;f(\omega + i\pi) = 2i\pi \sum_{k} \textrm{Res}\left(f,z_k\right)
\]
where $z_k$ are the poles of the function and $\textrm{Res}\left(f,z_k\right)$ stands for the residue of this function at that pole. For the function given at (\ref{eq: compl func 1}) we find that the poles are at $z_{1} = \frac{i\pi}{2}$ and $z_2 = z_1 + 2a$. Then, we find that 
\begin{equation}
\lim_{R\rightarrow \infty}\oint_\gamma dz\;f(z) = -i\frac{\pi}{3}\pi Q(a) =  2i\pi\sum_{k=1}^2\textrm{Res}\left(f,z_k\right)
\end{equation}
with
\begin{equation}
\sum_{k=1}^2\textrm{Res}\left(f,z_k\right) = 2a\csch{(2a)}-2\csch{(2a)}\coth{(2a)} + 4a\csch^3{(2a)}.
\end{equation}

The final result then reads
\begin{equation}
\label{eq: q}
Q(a) = 12a\csch{(2a)}-12\csch{(2a)}\coth{(2a)} + 24a\csch^3{(2a)}.
\end{equation}

It is important to remark that this result differs from those which are known in the literature for the same term in the collective coordinates effective lagrangian (usually also labeled as $Q$). In the seminal papers on this subject, \cite{belova} and also \cite{Sugiyama:1979mi}, this term is found to be divergent as $a\to 0$ while the above calculation shows that $\lim_{a\to 0}Q(a)=-2$, which makes the kinetic term of the $\xi$ coordinate in the effective lagrangin vanishes.

In the formulation of the effective lagrangian there is also another type of integral, for instance, in the case of the term labeled as $C$:
\begin{eqnarray*}
C(a) &=& \sqrt{\frac{3}{2}}\int_{-\infty}^{+\infty}dx\; \left(\tanh{(x+a)}\sech{(x+a)}-\tanh{(x-a)}\sech{(x-a)}\right)\left(\sech^2(x+a)+\sech^2(x-a)\right)\\
&=& \sqrt{\frac{3}{2}} \int_{-\infty}^{+\infty} dx\; \tanh{(x+a)}\sech{(x+a)}\sech^2{(x-a)}\\
&-&\sqrt{\frac{3}{2}} \int_{-\infty}^{+\infty} dx\; \tanh{(x-a)}\sech{(x-a)}\sech^2{(x+a)}
\end{eqnarray*}

Here the procedure is essentially the same, i.e., we consider the integral on the complex plane of the function $p(z) = f(z) - h(z)$ where instead of considering the complex functions $f$ and $h$ to be the complex extension of the real ones appearing in the integrands above, multiplied by $z$, we take it to be simply
\begin{equation}
f(z) = \tanh{(z)}\sech{(z)}\sech^2{(z-2a)}
\end{equation}
for the first of the two integrals above and 
\begin{equation}
h(z) = \tanh{(z-2a)}\sech{(z-2a)}\sech^2{(z)}
\end{equation}
for the second one. 

We take the path $\gamma$ to be the same as before once the poles of these functions are also $\frac{i\pi}{2}$ and $\frac{i\pi}{2}+2a$ and also here in the limit of $R \to \infty$ the integrals of these functions over the imaginary coordinate $\phi$ will give no contribution to the result. Then we have that
\begin{equation}
\lim_{R\rightarrow \infty}\oint dz\,p(z) = i\pi \sum_{k=1}^2 \textrm{Res}\left(p,z_k\right)
\end{equation}
with 
\[
\sum_{k=1}^2 \textrm{Res}\left(f,z_k\right) = -4i \csch^3(2a)\sinh^4(a) \quad \textrm{and}\quad \sum_{k=1}^2 \textrm{Res}\left(h,z_k\right) = 4i \csch^3(2a)\sinh^4(a).
\]

Finally, after simplification, the result reads
\begin{equation}
C\left(a\right)=\pi\sqrt{\frac{3}{2}}\tanh\left(a\right)\sech^2\left(a\right).
\end{equation}

\vspace{1cm}

{\bf Acknowledgements} 
GL would like to thank FAPES for the financial support under the EDITAL CNPq/FAPES N$^o$ 22/2018.
CFSP would like to thank FAPES for financial support under grant N$^\circ$ 98/2017.


 \end{document}